# Attilio Sacripanti*'

# Suwari•Seoi Safety: from children Dojo to High Level Competition (Biomechanical Part)

*IJF Academy Biomechanical Professor ' EJU Scientific Commission   Commissioner .
Tor Vergata University  Rome Italy

In this paper we face the problem of safety connected to a class of throws that are more often applied in every level of competitions.
Recalling the farsightedness of the founder of the judo, Jigoro Kano, on the safety of practitioners: establishment of the ukemi waza, and ban on the application of Yama Arashi, which has extended over time due to the sensitivity of the successors in the ban on kawazugake and Kani Basami.
This work is focused on the Tori safety in the application of the class of throwing applied with two knees on the mat, these techniques are very effective in competition, if right applied, but they seem to have a bad reputation, often connected to their premature application or to the chronic damage that can cause long-term agonistic activity, especially at a high level.
We must give credit to the Japanese Masters who seem to have, even in the texts, a greater sensitivity about the safety of the practitioners, in fact it is possible to read affirmation like the following ones.
*…. Numerous top judo competitors have had operations to remove damaged cartilages as a result too many drops with full force onto the knees. Currently (1990) in Japan, judo competitors under 16 (junior high school and under) are forbidden to attempt drop Seoi Nage on both knee in competition. With this rule, Japan hopes to maintain an overall high standard of basic judo technique, rather than give a handful of individuals to satisfaction of bringing home few medals. However, it must be said that as much as 80% of Seoi Nage seen in competition are this dropping version. [ Hidetoshi Nakanishi "Seoi Nage". Pag. 9-10 Publisher Ippon Books 1992 ISBN 0-9518455-4-3]*
 However, to our knowledge it has never been a scientific study on this subject in Japan and, to our great amazement, in Europe, only one, and in very recent times.
While the family of techniques that is applied with two knees on the ground has gone over time, increasing, without increasing the immediate traumas. this posed the problem if these techniques were really dangerous.
A group of researchers has decided to investigate in a comprehensive way, both on the immediate traumas, and on the chronic in the long run, and on the situations of competition. So, it was decided to develop the research on three complementary lines. Biomechanics of the technique aimed at the safety of Tori, from children in the dojo up to high-level competitions, a questionnaire that must have the greatest diffusion, seen the worldwide use of the technique, that provided information on the long term chronic trauma and an in-depth study of match analysis that highlight errors or safeguards, most common in competition.
With the purpose to clarify:
1) if the technique is dangerous, in its correct execution, for children and athletes.
2) if the technique is dangerous for the accumulation of traumas in the long run
3) if it is possible to find a possible safe form that could be gradually taught to children
4) if there are technical tools to make it more safe and effective for adult athletes
5) establish with a good degree of safety the age and time of teaching of these techniques.
6) build up a mathematical model of interaction between knees and Tatami
7) build up a safe way to train children
8) build up a training way to increase safety and effectiveness for Adult Athletes
The experimental part was developed with a thermal imaging camera that allowed to highlight the actual impact surfaces and therefore to trace the stress received by the knees, both for the children and for the athletes of the Italian National Team.
the results have made clear that, if well done and with a certified tatami, the technique has a negligible probability of immediate trauma.

With the worst stress on the Knees or Knees and Tibia ranging from 0,15 MPa till to 0,25 MPa. Equivalent, for a 12 cm thick knee in the sagittal plane, to a strain ranging between 0,4 mm till to 0,6 mm; compared to a physiological movement capacity between 5 mm to 10 mm.

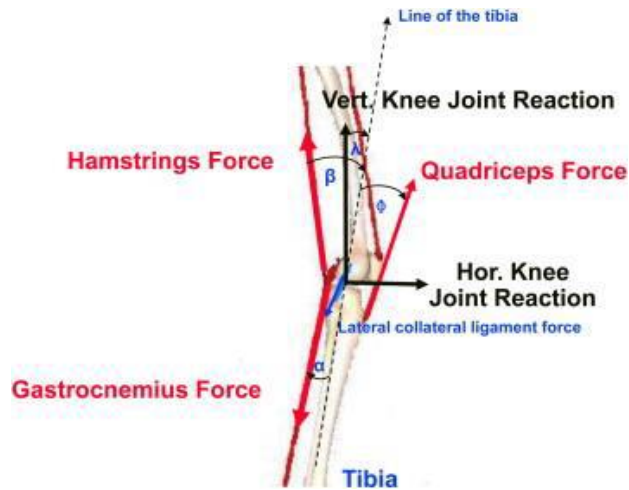

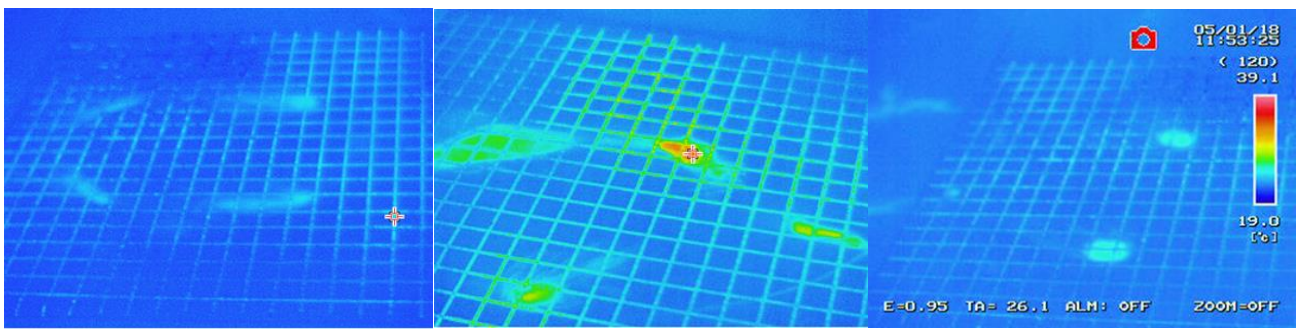

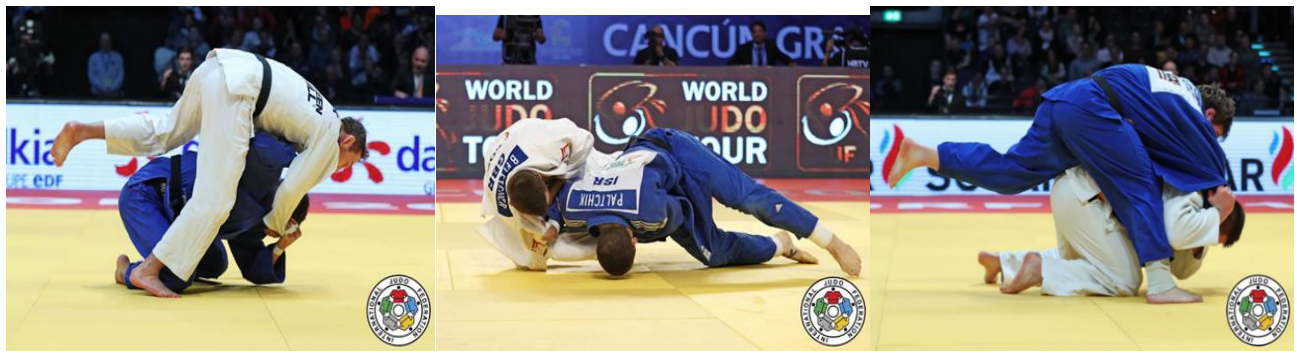

*The three main shapes of impact surfaces for Suwari Seoi family*

° the term Suwari means seated, often used in France and Italy was introduced here due to the lack of a Japanese denomination that does not distinguish between Seoi Otoshi and Suwari Seoi, as it is a technique of Lever Group, but with an arm of the leverage longer then with less application by Tori of the force to project Uke

# Attilio Sacripanti

# Suwari Seoi Safety: from children Dojo to high level competition

## Index

Attilio Sacripanti

*Suwari Seoi Safety : from children Dojo to high level competition*

*1 Introduction*

In this paper we face the problem of safety connected to a class of throws that are more often applied in every level of competitions.
As normal in such situation most discussions are available about right or danger to apply these techniques. Recalling the farsightedness of the founder of the judo, Jigoro Kano, on the safety of practitioners: establishment of the Ukemi Waza, and ban on the application of Yama Arashi, which has extended over time due to the sensitivity of the successors in the ban on Kawazugake and Kani Basami.
This complex work is focused on the Tori safety in the application of the class of throwing applied with two knees on the mat, both in seiza position with feet stretched and in Japanese style with feet pointed. These techniques are very effective in competition, if right applied, but they seem to have a bad reputation, often connected to their premature application or to the chronic damage that can cause long-term agonistic activity, especially at a high level.
We must give credit to the Japanese Masters who seem to have, even in the texts, a greater sensitivity about the safety of the practitioners, in fact it is possible to read affirmation like the following ones.
*…. dropping onto two knees, which has existed in judo as long as anyone can remember- has always been regarded as a lesser form, or even a bad form. Any young person who does it in a Japanese dojo will be severely criticized even now, for it is regarded as potentially damaging to their knees- which has turned out to be true. Numerous top judo competitors have had operations to remove damaged cartilages as a result too many drops with full force onto the knees. Currently (1990) in Japan, judo competitors under 16 (junior high school and under) are forbidden to attempt drop Seoi Nage on both knee in competition. With this rule, Japan hopes to maintain an overall high standard of basic judo technique, rather than give a handful of individuals to satisfaction of bringing home few medals. However, it must be said that as much as 80% of Seoi Nage seen in competition are this dropping version. [1]*
*….There has been a recent increase in incident where tori grapples in an extremely low posture, and enters deep inside uke, by dropping onto both knees for Seoi Nage or Seoi Otoshi. Seoi Nage with both knees dropped is, however, banned in the Kodokan referring rules and in junior judo… [2]*
Then it is important to connect rightly trauma developed in Judo with technical application, because the inverse solution will be to ban wrongly throwing techniques that are harmless.
We focus in this research on Seoi applied with two knees on the ground.
 How this technique is named?
In Japan Seoi Otoshi, both with one or two knees on the ground there is no difference at all, in English speaking countries Drop Seoi, in France, Italy and few other countries Suwari Seoi.
The names of Japanese throws born, as Kazuzo Kudo inform us: " *…Judo names fall into the following categories:*
1. *Names that describe the action*
2. *Names that employ the name of the part of the body used*
3. *Names that indicate the direction in which you throw your opponent*
4. *Names that describe the shape of the action takes*
5. *Names that describe the feeling of the technique.*

*Most frequently judo technique names will use the content of one or two of these categories [3]*

The name Drop Seoi, for the English countries, where do he was born?
Personal research led me to the 60s where I found for the first time the translation name:  Seoi Otoshi as Seoi drop, in the golden text of Koizumi [4], then during time seoi drop , changed in drop seoi to show the two knees variation of Seoi Otoshi .

There are differences between Seoi and Seoi Otoshi, with one or two knees?

For Japanese people we know that: *"It is important to discern the subtle differences between these two techniques….. Generally, tori should pull downwards with the body lowered when throwing with the knees dropped, but tori should load uke onto the back when throwing with knees not touching the mat. Therefore, we can define Seoi Otoshi as throwing with the knees dropped and seoi nage as throwing from a posture where the knees are not dropped."* [2]

In term of Biomechanics among Seoi, Seoi Otoshi and Seoi Otoshi with two knees on the mat, there is no difference as physical principle, it is  always the same application of the lever principle, the only changing aspect is the arm of the lever, that increases from Seoi to Seoi Otoshi to Suwari Seoi* ,  (* this name, as already underlined, is not in Japanese tradition but comes from France and Italian habit for  *sitting Seoi*, that for English speaking is called Drop Seoi).   for this reason, we face with three different energy consumptions and with three different stability and/or mobility situations for Tori.

Then basically we face the same physical principle with three different mechanical properties
On this basis we follow, to clarity of classification, the French denomination "Suwari Seoi" [5] specifically for the Seoi Otoshi variant with two knees on the mat.

From the technical point of view, rightly Japanese people when speak about Seoi throws speak about a family of throws similar but whit different arm position like Morote Seoi, Eri Seoi, Ganseki Otoshi , etc.

From the historical point of view about the naturalness and effectiveness of the two kneeling on the ground application, which we are reminded of by the two following figures, without wanting to go back to the examples of the wrestler's tomb in the Saqqara 3000 B.C. They are taken from the Greek-Roman world, in which Pancratius wrestlers and wrestlers are seen applying techniques of the "Suwari Seoi family".

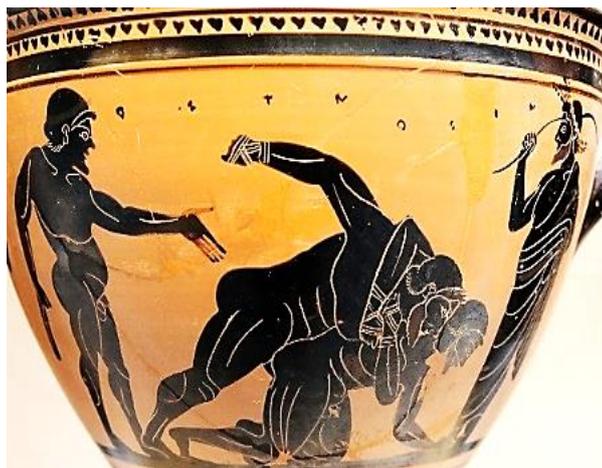

*Fig. 1  Pancratius wrestler applying a suwari seoi family throws*

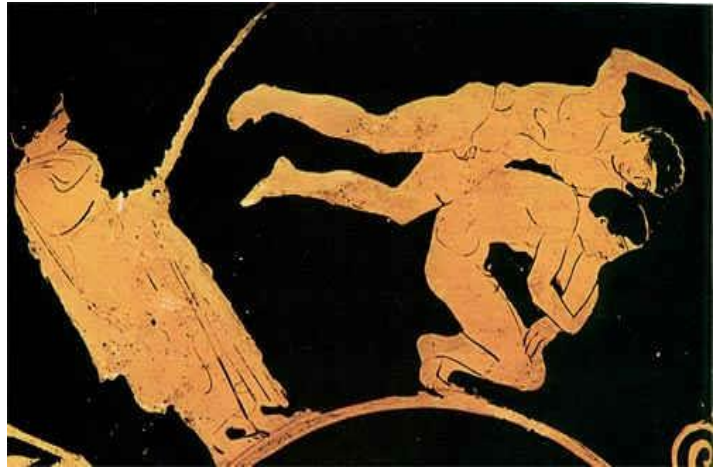

*Fig.2 wrestler applying a suwari seoi family throws*

## 2 Safety on Suwari Seoi family

In our situation "Knees' impact produced by Judo throwing techniques of the family of Suwari Seoi" the hazard is the condition that can cause injury to children applying these judo techniques or to adults used to apply them for long time life.

Obviously, the large range time of this study, needs a mixed approach in the research that it is not focalized on the Suwari Seoi poorly managed, but on the normal situation: children and adult that throw each other, with their Suwari Seoi throw.

Only few notations on the wrong application it is obvious that good teaching is a teacher duty, then big hazard are connected to teaching or to physically untrained subjects or by wrong technical methods.

The safety approach, needs to consider both the potential instant trauma, and the possible long-term trauma, very difficult to identify.

For the first kind of trauma, the medical literature [6] assures us that that from the position analyzed deep kneeling with feet pointed and seiza arriving position after jump, amazingly it could produce the posterior cruciate ligament (PCL) tear.

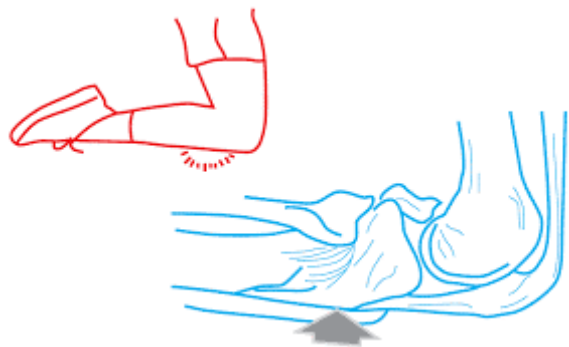

*Fig 3 Classical mechanism of Posterior Cruciate Ligament Tear*

The injuries to the posterior cruciate ligament (PCL) and posterolateral structures of the knee have received increased attention over the past decade; the result of this trauma is well known.[7,8]
The majority of patients who have isolated PCL tears can function with minimal disability; however, if the PCL is torn in combination with injury to the posterolateral structures, significant knee disability can result. The first mechanism of PCL tear is front car impact, the fall on the flexed knee with the foot in plantar flexion is the second most common mechanism of injury to the PCL .

When someone falls in this manner the tibial tubercle hits the ground, forcing the tibia posteriorly and tearing the PCL. A rotational mechanism associated with a varus or valgus stress may injure the PCL and collateral ligaments also. [9]

When we analyze the safety of Tori who apply a Suwari Seoi Family Throw, some thought must be focalized on the anatomical differences between male and female about the kinematics and kinetics of lower Kinetic Chains in Suwari Seoi Family techniques, as shown in the next figure.

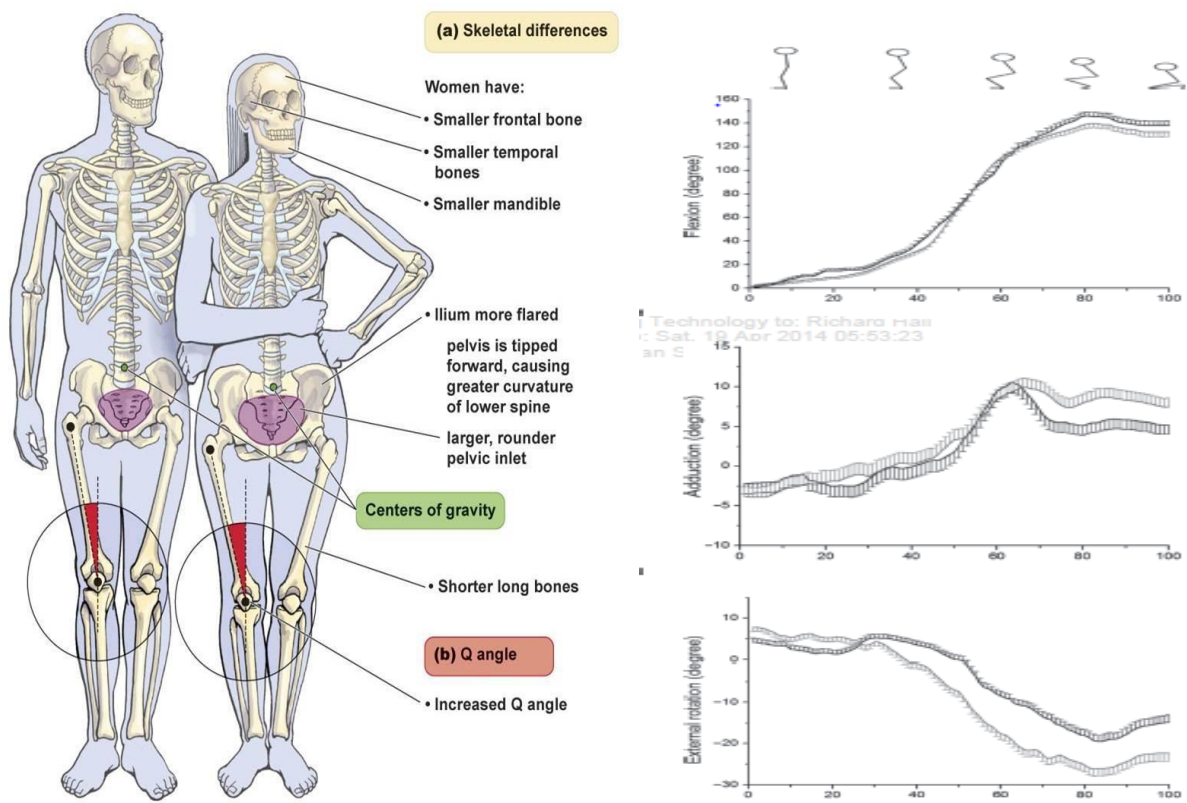

*Fig4 Differences between male and female knee*

Recent studies [10] on the kinematics of inferior chains during kneeling action showed interesting differences among male and female during deep kneeling action, which can be affect the safety of Suwari Seoi techniques specially in the frontal plane, connected with the ACL and PCL strain. Special care we must have for female athletes because the differences among males and females became apparent, in clear way, only when flexion angle was beyond 120°

These differences could be partially attributed to the difference in strength of the respective connective tissues (ligament, joint capsule, tendon, etc.) which were linked to the joints and may heavily influence the general joint's movements.

For example, it is well known that the posterolateral bundle of ACL played a crucial role in knee stabilization at high flexion angles (>120).

As Han and coworkers affirm: "… *The differences in secondary joint motions at high flexion angles may indicate varying ACL laxity between genders. The joint laxity of females could diminish joint proprioception, resulting in lower knee sensitivity to potentially injurious loads.*

*Meanwhile, ACL laxity would allow for a longer time for females to detect the joint motion. Due to a key role of ACL in controlling knee axial rotation, significantly higher knee rotations would be predicted in females".*

Female athletes appear to rely more on their quadriceps muscles in response to anterior translation, whereas male athletes relied more on their hamstring muscles. But high quadriceps loads would induce greater patellofemoral forces in females, which could be connected with the increased incidence of knee joint osteoarthritis in the female population used to sit in deep kneeling position.

Furthermore, the greater hip adduction, which was more apparent in females, could be another potential risk factor for ACL injury.
Generally speaking all the ligaments injuries and the ACL Injury could be divided in Direct and Indirect trauma [11] genesis:

**Direct trauma injury (rare):**
they occur as a consequence of a joint impact against an external body (contact / contrast) that occurs according to three main mechanisms:
• Through a direct trauma on the side wall of the knee that causes an external valgus-rotation;
• Through a direct trauma on the inner wall of the knee that causes a forced varus in internal rotation;
• Through a direct trauma to the back of the leg which causes anterior translation of the tibia ( the inverse mechanism figured in the fig 33 );

**Indirect trauma injury:**
 especially if the knee is in the " almost extension " position (about 20 ° of flexion). This is because the anteroposterior stability of the knee in these degrees is totally dependent on the ACL and the movement is poorly assisted by secondary stabilizers;

**The four main damaging mechanisms are:**
• **Valgus** - external rotation: the traumatic mechanism can occur during a fast movement, a deceleration followed by a change of direction: *possible in the complementary movements,*
• **Varus** - internal rotation: the traumatic mechanism can arise during the cutting maneuvers only performed by one knee during a complementary movement of turn plus elongation when Suwari technique is applied with the majority of weight is put on one only leg stretched and stopped ;
• **Hyperextension:** the traumatic mechanism can arise through a vacuum kick or a landing not properly stabilized, with hyperextension;
• **Hyperflexion:** the traumatic mechanism can arise as a consequence of a hyperflexion of the knee, followed by a powerful contraction of the quadriceps in an attempt to re-establish the upright position to throw suddenly the adversary on the back , normally this mechanism can arises to female athletes as Han and Coworkers remind us;

The probability of getting an Anterior Cruciate Ligament injury is, in women than in men, from four to six times greater.
The reason for this increased harmful frequency lies, as shortly showed before,  in the fact that women have some important anatomical and physiological differences.
• Less muscular strength, consequently also the lower stability control is lower;
• Flexor / extensor ratio more favorable to the extensors, therefore diminishes the defensive action of the ischi-crural muscle;
• Increased latency time, the action of the proprioceptive mechanism and the defensive mechanism of hamstring is slower;
• Increased flexibility and laxity, increased knee instability;
• ACL anatomically smaller, consequently withstands minor traumatic tensions;
• Larger pelvis to a greater external rotation of the tibia, are elements that favor the valgus of the knee and therefore increase the predisposition to the lesion in valgus-external rotation especially during the torsion in deep kneeling position during a complementary movement**.**
However, no PCL injuries, at our knowledge, were even diagnosed in judo as sudden trauma, connected to Suwari Seoi Family.
This means that, if well performed, the throws are safe also for female knees that are "more fragile" than male knees.
the critical nature of the complementary movements, on the safety of the knees, will be shown through a purely theoretical calculation in order to focus the attention of the teachers on a correct study of this part of the technique.
The theoretic results will be obtained in the par. 8. utilizing the equation {28}.

The only European study [12] and at our knowledge in the world, on the forces produced from a suwari seoi, shows interesting results, in the next two figures, we can see the layout of this research, and some quantitative results in term of forces suffered by Tori's knees, as a reaction to the fall due to the application of the suwari seoi.

Forces are evaluated in BW ( Body Wieght), in the next figure it is possible to see the contribution in N of the vertical force separated for components:  for Tori in Red more or less 400 N for Uke in Blue more or less 3500 N.

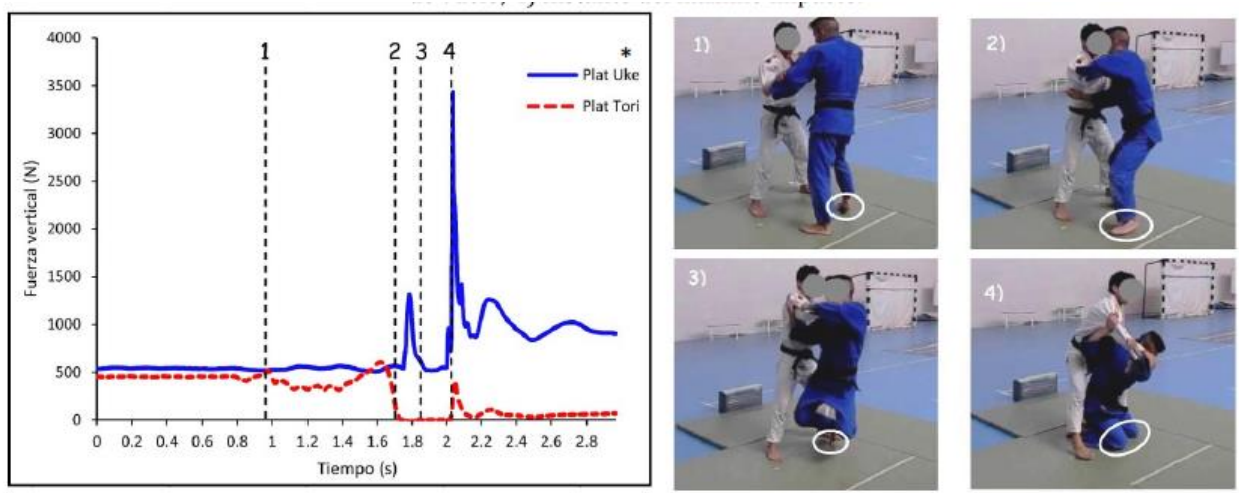

*Fig.5  Lay out and results in the Suwari Seoi analysis [12]*

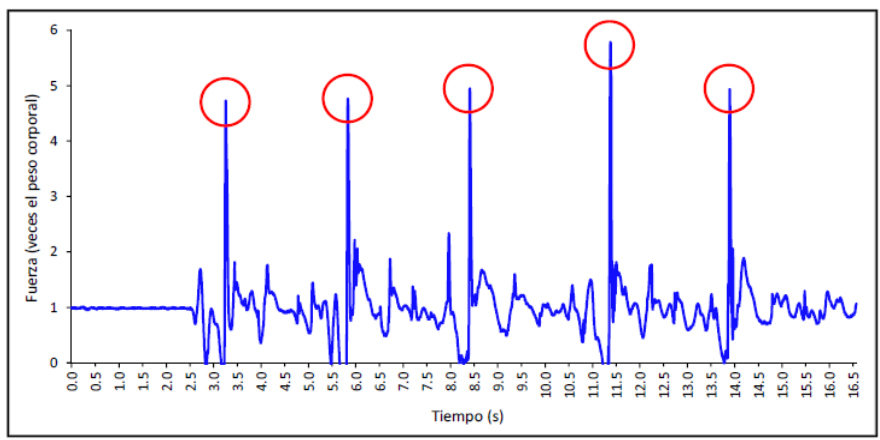

*Fig6   Peaks of reaction force suffered by Tori's knees [12]*

On the basis of this unique and specific work, some thesis were performed in Spain, one experimental was developed between two clubs, in Canary Island. [13]
One training children with Suwari Seoi the other one training children with Standing Seoi, to see differences in knees trauma between the two children groups. The total number was 80 children/cadets, (male and female) for each club they were analyzed for 10 months.
The results are shown in the next tables.

| Children | Club 1 | | | Club 2 | | |
|---|---|---|---|---|---|---|
| | *Suwari Seoi* | | | *Standing Seoi* | | |
| *Diagnostic* | *F* | *M* | *T* | *F* | *M* | *T* |
| *contusion with bruising and functional impotence right knee* | 9 | 6 | 15 | 4 | 2 | 6 |
| *Knee Sprain* | 1 | 1 | 2 | 1 | | 1 |
| *Traumatic Pain and functional impotence right knee* | 1 | | 1 | | | |

*Tab 1 Traumas and Suwari Seoi for Kids [13]*

| Cadets | Club 1 | | | Club 2 | | |
|---|---|---|---|---|---|---|
| | Suwari Seoi | | | Standing Seoi | | |
| Diagnostic | F | M | T | F | M | T |
| contusion with bruising and functional impotence right knee | 5 | 3 | 8 | 1 | 2 | 3 |
| Knee Spill | 1 | 1 | 2 | 1 | | 1 |
| Patellat tendinite Right knee | 1 | | 1 | | | |
| Dislocation of the patella right | | 1 | 1 | | | |

*Tab 2 Traumas and Suwari Seoi for Cadets [13]*

These results, all direct acute trauma, are very interesting because focalize our attention, in our opinion, probably more than the training mistakes, lacking teaching, and misuse of the throw, or not adequate tatami, than to the effective high danger of the techniques.

One notation is that every trauma is on the right knee, from that we can understand that the support on the knees was not equally balanced, but there was a preponderance of support on the right knee/leg, with consequent increase in pressure and actual risk of trauma.

These evaluations are supported not only by personal scientific experiences but also by the statistical analysis performed by the author, who demonstrated at the end of the work that there was no statistical difference between the results of these two techniques and that therefore the traumatic results, even though appearing different, were to be considered statistically equivalent.

Two interesting studies for adults, by questionnaire of connection between acute or chronic trauma and type of techniques. Here we can face with the lack of nomenclature, in fact Suwari seoi or Drop seoi etc, because are not defined as name, we do not know if they are presented as Seoi or Seoi Otoshi in the following studies nomenclatures.

The first one is a work on 78 Brazilian Athletes at regional level, [14] the second one 260 German Athletes at national level [15], the third one is Austrian Match Analysis on 69 competitions in Austrian National Championships 2014 and 2015 [16] which show not only the "safety" of the technique, here referred probably in the Japanese way as Seoi Otoshi, but also the knee very low danger connected.

**TABLE 3**
**Relationship injuries vs. type of stroke**

| Stroke | Gender M | Gender F | General total | Total (%) |
|---|---|---|---|---|
| Ippon seoi Nague | 12 | 4 | 16 | 23 |
| Tai otoshi | 10 | 4 | 14 | 22 |
| Uchi mata | 2 | 4 | 6 | 9 |
| Harai goshi | 1 | 3 | 4 | 6 |
| Briga de pegada | 3 | 0 | 3 | 3 |
| Chave de braço | 1 | 1 | 2 | 3 |
| O uchi gari | 1 | 0 | 1 | 2 |
| Sassae tsurikomi ashi | 0 | 1 | 1 | 2 |
| O goshi | 1 | 0 | 1 | 2 |
| Seoi otoshi | 1 | 0 | 1 | 2 |
| Hon-kesa-gatame | 1 | 0 | 1 | 2 |
| Koshi-guruma | 0 | 1 | 1 | 2 |
| Does not remember | 8 | 5 | 13 | 22 |
| Total | 41 | 23 | 64 | 100 |

*Tab 3. Brazilian Athletes traumas throws connection Questionnaire [14]*

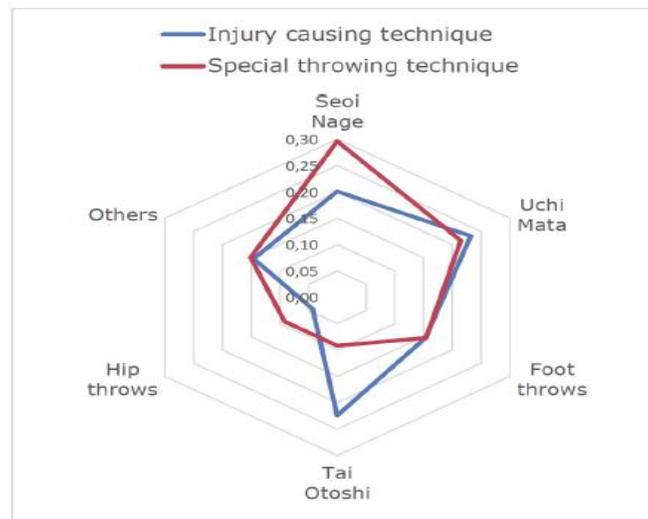

*Diag 1 special and injury causing ttrhows*

|  | Uchi mata | Seoi nage | Tai otoshi | Foot throws |
|---|---|---|---|---|
| Muscle injury | 0.21 | 0.20 | 0.26 | 0.29 |
| Contusion | 0.38 | 0.70 | 0.43 | 0.21 |
| Meniscus | 0.50 | 0.30 | 0.52 | 0.43 |
| ACL | 0.46 | 0.15 | 0.26 | 0.07 |
| PCL | 0.00 | 0.05 | 0.04 | 0.07 |
| Patella | 0.04 | 0.15 | 0.17 | 0.36 |
| MCL | 0.25 | 0.10 | 0.22 | 0.21 |
| LCL | 0.13 | 0.10 | 0.13 | 0.43 |
| Others | 0.46 | 0.45 | 0.35 | 0.00 |

*Tab 4- German Athletes Traumas/ Throws connection and kind of traumas, Questionnaire [15]*

| INJURY CAUSED TECHNIQUE GROUP |  |
|---|---|
| grip fight | 80,0% |
| sacrifice throwing techniques | 6,67% |
| joint locking techniques | 6,67% |
| leg throwing techniques | 6,67% |
| TOTAL | 100,0% |

| POSTURE AT INJURY |  |
|---|---|
| standing posture | 93,3% |
| ground posture | 6,7% |
| TOTAL | 100,0% |

*Tab. 5-6 Austrian Athletes techniques and posture connected to injuries [16]*

| INJURED BODYPARTS | male sex | female sex |
|---|---|---|
| face | 9,1% | 0% |
| nose | 27,3% | 25% |
| mouth | 9,1% | 0% |
| ellbow | 9,1% | 0% |
| hand & fingers | 27,3% | 50% |
| thorax / abdomen | 9,1% | 0% |
| knee | 0% | 25% |
| toes | 9,1% | 0% |
| TOTAL | 100% | 100% |

*Tab. 7    % of Injuries placement in Austrian Championships 2014-2015 [16]*

## 3 The aim of this research.

Because we are corroborated by the absence in the scientific literature of immediate damage to the PCL connected to Suwari Seoi techniques, we started from the hypothesis that the two knees family of throwing techniques, defined for short. "Suwari Seoi" are not dangerous if properly executed on a good tatami, like all other throwing techniques accepted in Judo; the research group will analyze the correctness of the hypothesis, analyzing from the safety point of view, all sides of the problem.
We assume that the real problem could be for the long-term trauma, the overuse, and if present for the acute trauma produced or by the wrong application of the throw or by the follow-up of the technique that is based on complex rotational movements starting from the kneeling position.
To have data, for the long-term trauma, the only way is to prepare a questionnaire and spread it around. This is the questionnaire reason. one focus of our research.
The other step is to demonstrate that (in safety term) Suwari Seoi it is not dangerous, obviously strange accident are every time present, for dynamical reasons, and we cannot prevent it!
But if we are able both to demonstrate the safety of throw and his right biomechanical principle of application, then with a correct teaching of the safe application in training we can decrease the possibility that accident for wrong use of the throw can happens.
This is the second focus of our research
How it is possible to demonstrate that?
The research protocol is made with specific attention to the motion and impact of Tori's knees, to prevent acute trauma.
We have not Force platform, because simple physics let us to know both the impact force and velocity, then we ask for each subject two specific trials with three subcases, six performances of throw.
The first trial is with knees falling down with both feet pointed touching the tatami, then the mechanics of this way of application it is a fall with control, specifically a mechanical paradox ( see biomechanics paragraph 6 ), and the area of percussion is smaller than the other way to touch the Tatami, only the tips of pointed feet and small tibial area, we take time of fall's trajectory, to evaluate the mean speed of fall.
The second trial is the other way to apply Suwari Seoi in competition, jumping with the feet stretched out not touching at all the tatami , in that case the mechanics of the action is a free fall and the contact area is larger, long tibial surface projection and back of the feet.
We take the contact surface by the thermal image on the tatami, this area is a % of total body surface area BSA that is evaluated by Du Bois and Du Bois formula. [17]
For both trials the safety approach is translated as a fall from the knees' height, measured before the trial.
Third focus of the research is to develop a mathematical model of the impact, the mathematical model is a knee (very simplified model) that collides with a Tatami that is built with viscoelastic substance. the model will be understandable and will show the differences among the knee impact starting from the previous feet positions.
The mathematical model is built because it is possible to change values at some specific parameters like hardness or softness of the mat, falling velocity, height of fall, and see the effect produced without any other use of children or athletes
Fourth focus of the research is the confrontation between children and adult in term of styles velocity, and other aspect like safety, etc.
Fifth focus is the analysis in competition to know how many times the throw is performed, in which fashion ( feet pointed or stretched ), how it is effective. and try to classify the further movements after the knees impact used to perfect the final part of the throw, the so called "Complementary Tactical Tools" ( CTT). [18]

From all that we will try to single out the best biomechanics form to apply the technique in such way we are able to give some suggestion for children training in safety and some to optimize the adult way to throw.

For high level we will try to give some global indication about how to perform this technique in effective but safer way.

*4 Biomechanics of Suwari Seoi family*

As it is well understood the mechanics of Suwari family is a lever mechanics, with maximum arm, but there is a more subtle difference between the applications studied A ) with feet pointed ,and B) with feet stretched, when we analyze the motion and the fall of the knees: in the first example we face with a mechanical paradox, and in the second one with a pure free fall example.

This means that equation that explain the motion dynamics of legs and knees are different also from the theoretical point of view. In the first case A) under some specific conditions the knees vertical downward acceleration can be higher than g, while in the second case B) the acceleration will be always g but the conservation of energy assures us that the impact force of the knee will be higher than the athlete's weight because inside there is the restitution of jumping up energy before utilized for the starting jump of the throw movement. [19]

A) In the situation of the pointed feet is very interesting, the net moment on the leg is
$\tau = k\, Lmg \cos \varphi$   {1}
With L= length of leg, k= <1 (variable location of CM) , m =body mass,  g =gravity acceleration,

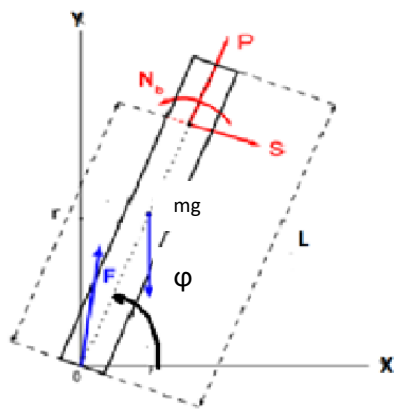

*Fig 7 Model representation of lover leg*

The equation of the dynamic motion of the leg is: $\tau = I\alpha = I\dfrac{d\omega}{dt} = I\dfrac{d^2\varphi}{dt^2} = kLmg \cos\varphi$ {2}

The inertial momentum is $I = \beta mL^2$ {3} from the equation 1.2 and 1.3 it is possible to evaluate the angular acceleration α.

that is : $\alpha = \dfrac{kg}{\beta L}\cos\varphi$  {4}

if we define the constant $\omega_c^2 = \dfrac{kg}{\beta L}$  {5}

it is possible by integration to obtain the angular velocity ω :

$$\omega^2(t) = 2\omega_c^2 \left(\sin\varphi_0 - \sin\varphi\right) + \omega_0^2 \Rightarrow \left(\dfrac{d\varphi(t)}{dt}\right)^2 = 2\omega_c^2 \left(\sin\varphi_0 - \sin\varphi\right) + \omega_0^2 \quad \{6\}$$

The differential equation gives us the variation in time of the angle by the solution of elliptical integral. It is also very important to analyze the position of the total CM of body, because normally if the mass is put before the collision center of the leg it can accelerate the vertical velocity of knee.

In fact, the acceleration of the knee is:

$a_T = \alpha L$  {7}   and the normal component of this acceleration to the mat is:

$$a_n = a_T \cos\varphi = \dfrac{kg}{\beta}\cos^2\varphi \Rightarrow \dfrac{2}{3}g\cos^2\varphi \quad \{8\}$$

In some condition of throws body weight, during the controlled falling in deep kneeling with feet pointed, is moved near the feet position

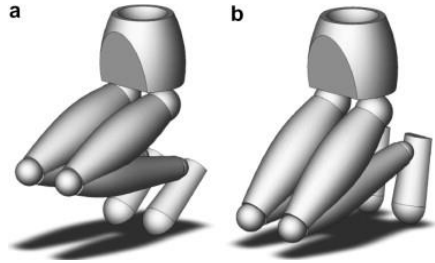

*Fig 8. way to fall with feet pointed*

In such condition if we call M the body mass and m the mass of the leg calf ( % of M) the tangential acceleration of knee in the horizontal position is [20] :

$$a_T = \dfrac{3}{2}gL\left[\dfrac{Ml + mL}{Ml^2 + mL^2}\right] \quad \{9\}$$

we can analyze the two limit situations: Ml=mL  and M>>m; and easily we obtain:

$$a_T \equiv a_n \simeq 3g \quad \{10\}$$

$$a_T \equiv a_n \simeq \dfrac{3L}{2l}g \quad \{11\}$$

From the previous two equations it is easy to evaluate the range of falling velocity of the knee in these limit situations and we see that it ranges among 2.6 m/s < v < 6 m/s.

In term of safety we are now able to evaluate the maximum impact force of knee on the tatami.

Other important information from this theoretical approach is the evaluation of time to hit the tatami, calculations are difficult but readers can find it in references [a,b,c] we give only the final result useful for our safety analysis in term of ratio between the time of free fall divided by the time of knee point , if the ratio is less than 1 the condition of faster than g is satisfied.

$$\frac{T_0}{T_1} = \sqrt{\frac{2k \sin \varphi_0}{\beta I_1^2}} \quad \{12\} \quad \text{in which the time of free fall is} \quad T_0 = \sqrt{\frac{2L}{g} \sin \varphi_0} \quad \{13\}$$

Moreover, with good mechanical approximation it is possible, considering this situation as model of a falling chimney, to obtain rough data about internal stress of the knee following and adapting the results of Varieschi and Kamya [21] at our specific situation. For the leg calf that progressively bends in some situation potentially it is under the combined actions of a longitudinal force P and a bending moment N, the stress at knee considering the bone leg composed by one homogenous piece is in non-dimensional form:

$$\frac{\pi \rho^2}{mg} \sigma_k = \frac{1}{2}\left(1 - \frac{r}{L}\right)\left[\left(5 + 3\frac{r}{L}\right)\cos\varphi - 3\left(1 + \frac{r}{L}\right)\right] + \frac{3}{2}\frac{L}{2\rho}\frac{r}{L}\left(1 + \frac{r}{L}\right)^2 \sin\varphi \quad \{14\}$$

This equation depends from the L/2ρ, with L= length of calf and 2 ρ = bone diameter, in the human body case, the dimension is about 66 times, this means that the second term of this equation is enhanced by the previous ratio and the term depending by the bending moment plays a more important role in the safety of athletes' knees.

From the other side the shear stress, can be the other leading cause of rupture. It is easily seen that, for any specific angle, the magnitude of the shear force has an absolute positive maximum and usually originate near the ankle, meaning that large shear forces, can affect the ankle joint that serves as a pivot in the throwing action.

B) In this other situation, athlete jump and land on the tatami with feet stretched, landing in "seiza" position, the mechanics under this throwing style is the simple mechanics of human body free fall.

In term of safety we must consider that the impact energy is function of the height of the jumping. In equations we can write remembering the conservation of mechanical energy:

$$\frac{1}{2}mv^2 = mgh \quad \{15\}$$

The applied load on the knees and legs during free landing varies according to the height, ground softness, joint flexion, landing positions and direction. The magnitude of the load obviously increases as height of jump increments, whereas the impact time decreases. Considering that no significant differences can be found among the falling periods from various heights, because athletes and children are used to no jump high but they jump as slipping between the Uke legs.

The impact time (impact duration) was quite short.

Therefore, an average time of 0.04 s was applied in the analysis. The landing time was evaluated for each trial.

The velocity of each subject during free-fall in term of safety was evaluated starting from the knees' height, using the following well known elementary equations:

$$\frac{1}{2}mv^2 = mgh \Rightarrow v_f = \sqrt{2gh} \quad \{16\}$$

whereas the average theoretical time to fall was calculated by the following relation:

$$t = \sqrt{\frac{2h}{g}} \quad \{17\}$$

The average impact time derived during the experimental test was, as previously expressed, equal to 0.04 s. This time allows that the impact load to be identified using the following equation:

$$\int_0^t F dt = \int_0^{v_f} m dv \quad \{18\}$$

Many studies are performed on biomechanics of judo throwing techniques, but in our knowledge very few on the Suwari Seoi, mainly in France and Japan, whereas from Spain comes the highest number of theses and papers that advise against the use of this technique for children.

The main aim of the biomechanical studies is to understand "how things work ", for this reason the studies' approach are very similar because the things to understand are the same. From this point of view the French study [22,23] was focalized on the "principles" of effectiveness of the execution of Suwari Seoi, by mechanical measurements of the movement. To singling out the fundamental skill of the techniques.

In the following figures we can see: the motion capture of Suwari Seoi, Fig 7, the velocity of CM during 5 throws from 5 World champions Fig.8 , global rotation around z axis, Fig.9 global rotation around y axis.Fig.10 from this figure we can understand the different specific styles of throwing action in Suwari Seoi

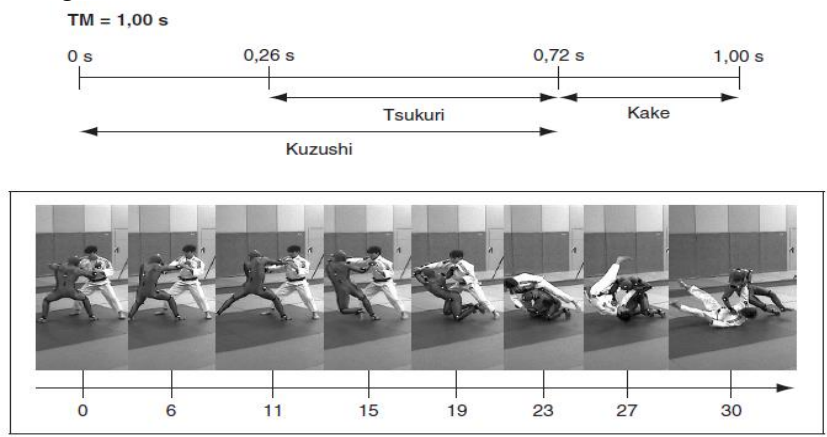

*Fig 9 Motion Analysis of Suwari Seoi [21]*

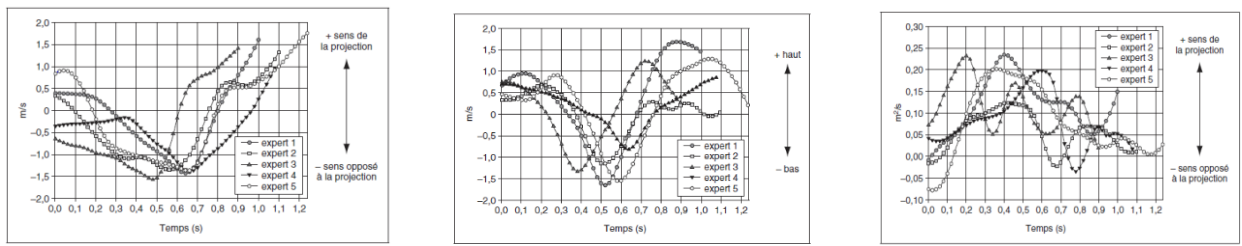

*Figg. 9,10,11   Cm and Global rotations around axis [21]*

Ishi e coworkers performed a very long study on Seoi  [ 24,25,26,27,28]  from 2004 till 2016 also comparative between Japanese champions and students to understand the effectiveness of Seoi action, studying the front turn movement, the kinetics of legs, CM motion, angular movements and so on.

In the following figures it is possible to see some results of these researches.

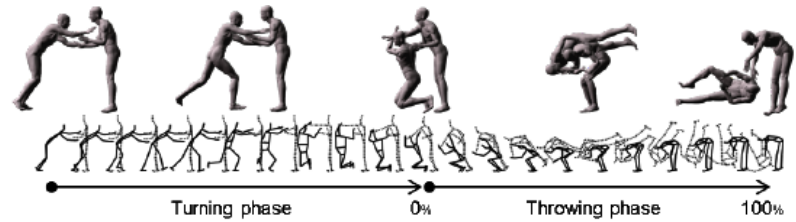

*Fig.12 Definition phases in Ishi experiment*

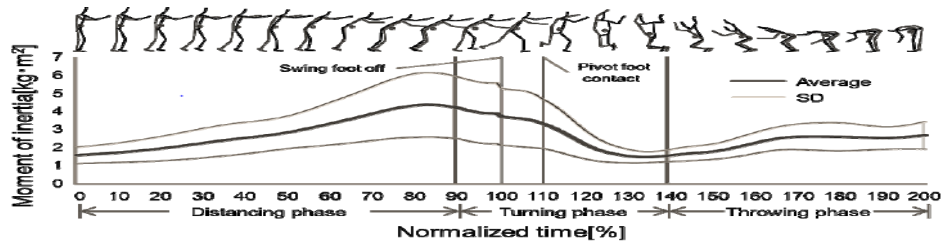

*Fig 13 Time evolution of Moment of Inertia [24]*

Table I. Difference in the kinematic and kinetic variables between elite and students athletes.

| | Elite A | Elite B | Elite C | Students (M ± SD, n=7) |
|---|---|---|---|---|
| **Peak angular velocity (rad/s)** | | | | |
| Knee of pivot leg | 6.4 | 7.5 | 7.5 | 4.0 ± 1.8 |
| Knee of swing leg | 7.0 | 8.2 | 8.6 | 4.7 ± 3.0 |
| Hip of pivot leg | -5.7 | -7.6 | -6.3 | -3.6 ± 1.2 |
| Hip of swing leg | -9.0 | -10.7 | -9.1 | -4.9 ± 4.5 |
| **Peak extension torque (Nm/kg)** | | | | |
| Hip of swing leg | 2.42 | 4.80 | 3.53 | 1.60 ± 0.83 |
| **Peak flexion torque** | | | | |
| Knee of swing leg | -1.14 | -1.49 | -1.19 | -0.50 ± 0.75 |
| **Peak positive torque power (W/kg)** | | | | |
| Hip of swing leg | 7.99 | 15.73 | 9.19 | 3.12 ± 3.46 |
| **Peak negative torque power (W/kg)** | | | | |
| Knee of swing leg | -6.96 | -9.69 | -10.20 | -1.95 ± 4.10 |
| Hip of swing leg | -5.47 | -16.98 | -9.38 | -1.72 ± 2.52 |

*Tab. 8 Comparative analysis of Seoi kinetic and kinematics variables [26]*

About the standing seoi technique a lot of studies have ben performed around the world, starting from the historical Japanese work of Ikai and Matsumoto 1958 [29], most it is now known but about his biomechanics physiology and so on [30,31,32,33,34,35, 36,37,38,39,40,41,42,43]. Instead about Biomechanics of Suwari only the French study was performed and in term of safety only the present one and the cited Spanish study [12] were developed.

## 5 Tatami Material Science and Thermodynamics.

Material science is essential part of this research, because safety outcome depends both from the Tatami material and quality [44] . In this research we analyzed one Tatami built by polyurethane foam and soft polyurethane covered by pvc and Approved by IJF, with thickness of 4 cm, and overall density 240 kg/m³. tensile strength 2480 N/5 cm, theoretical Force reduction ≈ 25%-40%

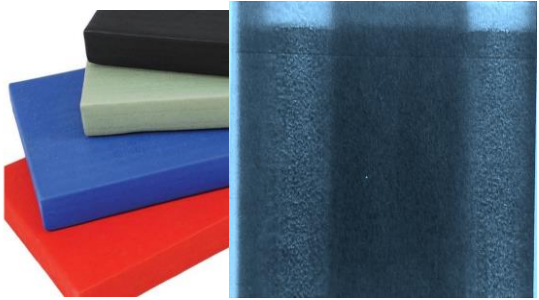

*Fig 14 Tatami and vertical constituents section*

PU invented by Bayer in Germany around 1937, have a history of slightly more than 75 years. They have become one of the most dynamic groups of polymers. Their use covers practically all fields of polymer application: foams, elastomers, thermoplastics, thermo-rigids, adhesives, coatings, sealants, and fibers.

PU are obtained by the reaction of an oligomeric polyol [low-molecular weight (MW) polymer with terminal hydroxyl groups] and a polyisocyanate.

The structure of the oligomeric polyol used for PU manufacture has a very profound effect on the properties of the resulting polymer, as assure us Ionescu in his encyclopedic work [45]. The tatami analyzed was built by three layers first layer PVC, second Polyurethane foam, and third Polyurethane semi rigid.

The foam is important but his mechanical evolution is quite complex.

The response of foam gets stiffer with increase in strain rate, and densification (lockup) occurs well below the strains at which lockup occurs for foam deformed at quasi-static strain rates. Consequently, the energy absorption characteristics of foam are altered with change in strain rate. [46]

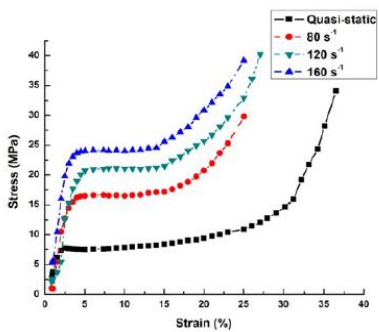 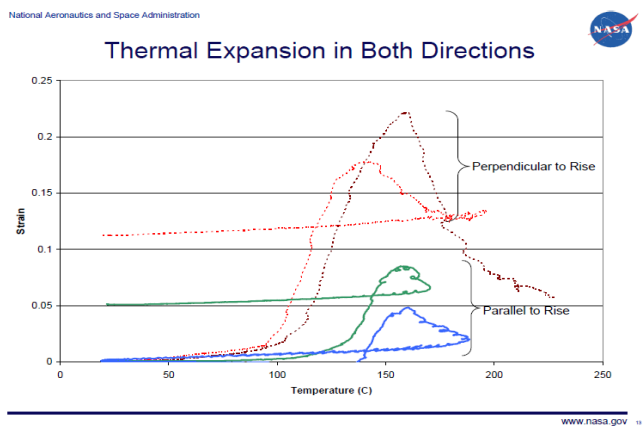

*Diag 2  Stress strain foam curves[46]   Diagg 3 Polyurethane Foam thermal expansion [46]*

Also very complex is the thermal behavior of polyurethane foam, as NASA researchers shown in some very interesting works [47].

Since the foam is not a material, but a structure, the modeling of the expansion is complex. It is also complicated by the anisotropy of the material. During the spraying and foaming process, the cells become elongated in the rise direction and this imparts different properties in the rise direction than in the transverse directions.

However we are much more interested than expansion, in his compression and related thermo-dynamical effects.

If the compression produced by children's body is fast, the situation can be approximated in thermo-dynamical terms to an adiabatic transformation.

This specific transformation was named by Viecheslav Sychev in his book "Complex thermodynamic System" [48] Elasto-caloric Effect.

*5.1 Elasto-caloric Effect*

When a body falls, on the tatami, after a Judo throw; the impact produces one adiabatic compression of the tatami, the impact energy will partially have absorbed and one of the main effect, that changes the mechanical energy into heat, is the *Elasto-caloric Effect*.
The induced variation of temperature is expressed by the following easy calculation:

$$T = T_0 + \int_0^{\Psi} \left(\frac{\partial T}{\partial \Psi}\right)_{S,P} \partial \Psi \quad \{19\}$$

To solve the kernel of the integral we can use the Maxwell equation

$$\left(\frac{\partial T}{\partial \Psi}\right)_{S,P} = -\left(\frac{\partial l}{\partial S}\right)_{\Psi,P} = -\left(\frac{\partial l}{\partial T}\right)_{\Psi,P} \left(\frac{\partial T}{\partial S}\right)_{\Psi,P} \{20\}$$

And after few simple calculations we have the following final Relationship:

$$T = T_0 - \frac{\alpha_l \overline{T}}{c_p \rho} \Psi \Rightarrow \Delta T = -\frac{\alpha_l \overline{T}}{c_p \rho} \Psi \quad \{21\}$$

When the Tatami is compressed the stress $\Psi$ is negative and the Tatami temperature increases, absorbing energy.
The previous final relationship {21}, remembering the Hookean Elastic Equation can be changed as:

$$T = T_0 - \frac{\alpha_l \overline{T}}{c_p \rho} \Psi \Rightarrow \Delta T = -\frac{\alpha_l \overline{T}}{c_p \rho} \Psi = \frac{\alpha_l \overline{T} E \Delta \varepsilon}{c_p \rho} \quad \{22\}$$

To have a first indicative order of magnitude in our research, very simple "theoretical" evaluation, assures, that with Polyurethane Foam as Tatami material IJF Licensed, with **density 244 Kg/m³**, with energy absorption, **around 25%- 35%**, supposed: *Theor. Compr* ≈ 2 mm, temperature will have a "theoretical" increase of : **296.15<T(°K) <297.0** or in Celsius **23 < T(°C) <23.8**

Mechanics and Elasto-caloric effect are connected by means of Strain that Produces Tatami Compression by Hook law.

$$\Psi = \frac{F}{A} = E\Delta\varepsilon \Rightarrow \Delta\varepsilon = \frac{\Delta l}{l} \quad \{23\}$$

Compression is produced by Tori knee falling down in the two ways analyzed and part of the Strain, after energy absorption, is returned to Tori body for the Action Reaction Principle. In formulas:

$$\Psi' = -e\Psi \quad \{24\}$$

In which ($e < 1$) is similar to the restitution coefficient and depends from the Tatami Material.

Remembering that:

$$\Psi' = -e\frac{F}{A} \quad \{25\}$$

A in the equation {25} is the surface that in our current research corresponds to the actual impact surface of Tori carrying out the Suwari Seoi, which in our two experimental trials examined corresponds to two basic positions:
- Pointed feet: summation of tibial protuberance area and tip of the toes. Fig A
- Stretched feet: total area of the contact of the tibia plus that of the back of the foot. Fig B

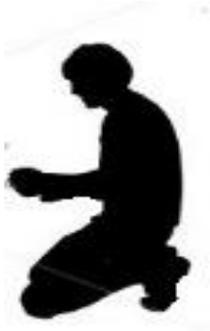   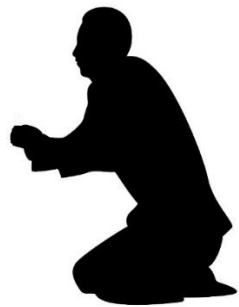

*Fig. 15 Deep Kneeling feet pointed*         *Fig. 16  Kneeling with feet stretched " Seiza" position*

## 6. The throw

The application of suwari seoi family in competition needs some specific biomechanical situation. It is well known that both Seoi Nage and Suwari Seoi Nage are among the most utilized and successful judo throws in the world. As it easy to see in the next tables: Tab 9,10 .

| FRA | | JPN | | URSS | | Autres | |
|---|---|---|---|---|---|---|---|
| Uchi-mata | 25,5 % | Uchi-mata | 15,8 % | Uchi-mata | 11,4 % | Suwari-seoi-nage | 13,8 % |
| O-uchi-gari | 11 % | Suwari-seoi-nage | 13,3 % | Seoi-nage, Kata-guruma | 9,6 % | Uchi-mata | 13,4 % |
| O-soto-gari | 7,7 % | Ko-uchi-gari | 10,7 % | Suwari-seoi-nage | 7 % | O-uchi-gari | 8,7 % |
| Sode-tsuri-komi-goshi | 7,4 % | O-uchi-gari | 9 % | Kuchiki-daoshi | 7,7 % | Ko-uchi-gari | 7,4 % |
| Ko-uchi-gari | 7,4 % | O-soto-gari | 7,3 % | O-soto-gari | 7,4 % | Seoi-nage, Kata-guruma | 7,4 % |
| Kuchiki-daoshi | 7,4 % | Tomoe-nage | 6 % | Sode-tsuri-komi-goshi | 7 % | Hara-goshi | 7 % |
| Suwari-seoi-nage | 5,3 % | Seoi-nage, Kata-guruma | 5,1 % | Tai-otoshi | 7 % | | |

*Tab. 9 % of utilization of judo throws by athletes from France Japan and Russia [5]*

| | FRA | JPN | URSS | Autres |
|---|---|---|---|---|
| 1 | Uchi-mata | Suwari-seoi-nage | Seoi-nage, Kata-guruma | O-soto-gari |
| 2 | Kuchiki-daoshi | Seoi-nage, Kata-guruma | Uchi-mata | Suwari-seoi-nage |
| 3 | Seoi-nage, Kata-guruma | Sode-tsuri-komi-goshi | Ura-nage | Sode-suri-komi-goshi |
| 4 | O-uchi-gari | Ko-soto-gari-gaké | O-uchi-gari | O-uchi-gari |
| 5 | O-soto-gari | Tomoe-nage | Sode-tsuri-komi-goshi | Tai-otoshi |
| 6 | Hiza, Sasae | O-soto-gari | Maki-komi | Tomoe-nage |

*Tab 10 Throws effectiveness for athletes from France, Japan and Russia [5]*

Biomechanical analysis shows other interesting findings about Seoi throws. Generally speaking is impossible to perform Seoi techniques without unbalance, carefully all the Seoi techniques that are kuzushi (unbalance) dependent, they need also to stop for a moment the adversary motion to be carried out. Then the Seoi family throws are complex movements that need the highest coordination and timing in motion, specially to obtain the optimal tsukuri position and overcome the strength defensive opposition of Ukes' grips ( Kumikata), these are, in general, with the specific mechanics of throw, the reasons of the high energy needs for Tori. But the maximum of Kano: the best use of energy and the maximum effectiveness with the least possible effort, is unconsciously applied in high-level competitions by all athletes and therefore the family of Suwari Seoi is very often applied with great success. Having, for Tori, a series of undeniable advantages over the standing Seoi family.

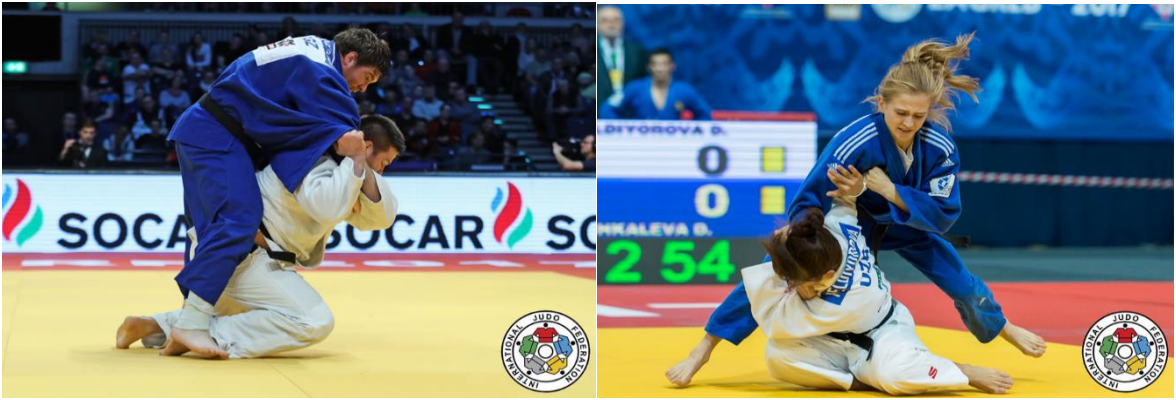

*Fig17 Male and Female version of Suwari Seoi Family*

*Tori's Advantages toward Classic Seoi*

1) *Energetically less expensive, (the arm of the lever is longer).*
2) *Less important Kuzushi ( the drop down movement acts as kuzushi)*
3) *Easier Tsukuri positioning (it is Uke's body that fits Tori's body and not the inverse).*
4) *Useful in overcoming Uke's grips (arms are less able to stop up and down movement, than push/pull actions).*
5) *Useful in increasing rotational speed*
6) *Most difficult Uke's Avoidance vs drop action*

However, these previous multiple and undeniable advantages obtained that simplify the action of Tori are, on the other hand, paid with two important complications.

*Tori Disadvantages*

1) **Less control over Uké body defensive direction**
2) **Drastic increase in stability and decrease in mobility**

In high-level competitions, due to athletes' defensive acrobatic skills, it is difficult for Tori, with his knees touching the tatami, to manage the direction of throwing forces according to the defensive movements of Uke, when the throw is not perfect.

This is the reason why it becomes necessary to use additional movements aimed at perfecting the action in order to obtain an effective result. So, dynamicity of a throwing action, high defensive skill and lesser mobility are the three main reasons to apply tactical support tools (specific movements) to refine action in competition, because during the fall down Tori have few control of Uke movement, and the end of trajectory is a situation of multiple direction choice for Tori and this is also the main difficulty to manage this throw in effective way.

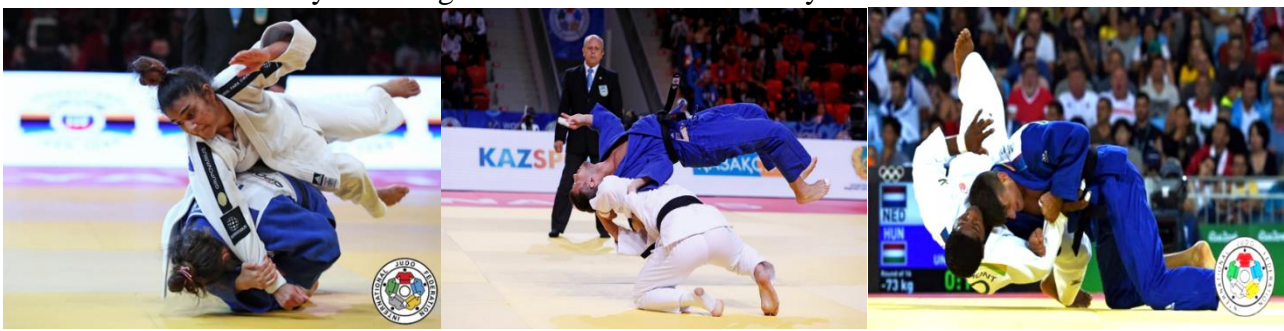

*Fig 18 Esoteric, but effective, Variation of Suwari Seoi Family*

## 7. *The complementary movements (tactical tools)*

When athletes perform Suwari Soi in competition, because it is very difficult for Tori to control the defensive body direction of Uke, [50] arises the necessity to perfect the final part of throw using some complex movements usually called *"tactical tools"[51]*.

Starting from the two body's positions (a) Seiza and (b) Kneeling, that represent the basic arrival position of Tori body after the drop down, all tactical tools start with a leg extension in different directions (mainly: up, forward, diagonal) connected with some complementary body movements, like: torque, push, body's rotation or flexion, helped by some specific arms movements

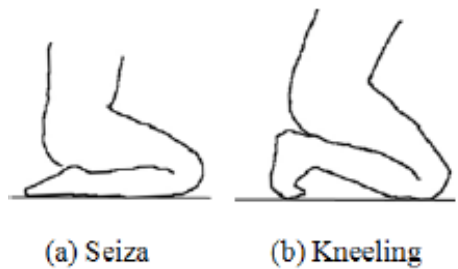

*Fig 19 feet positions*

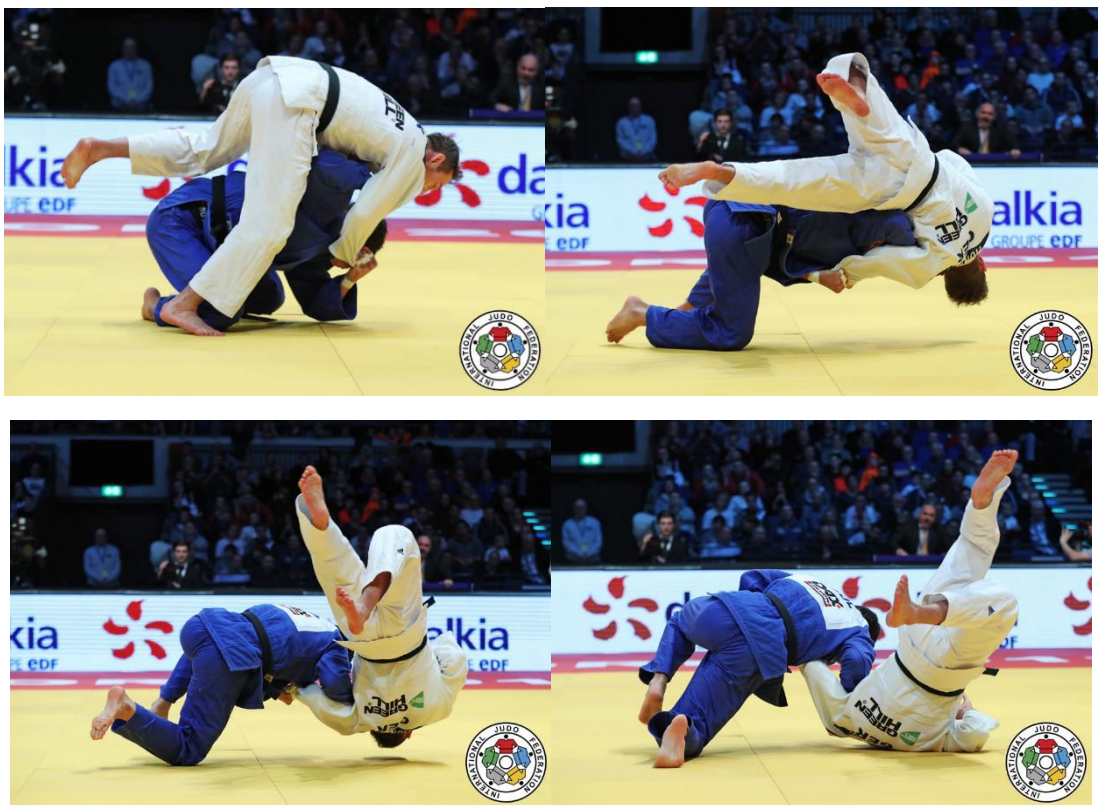

*Figg.20-23 complementary movements to obtain full point ( Ippon)*

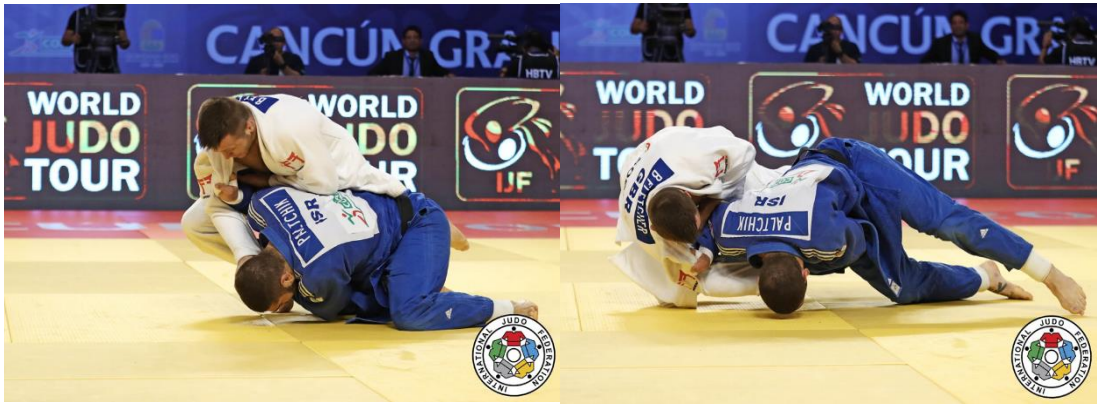
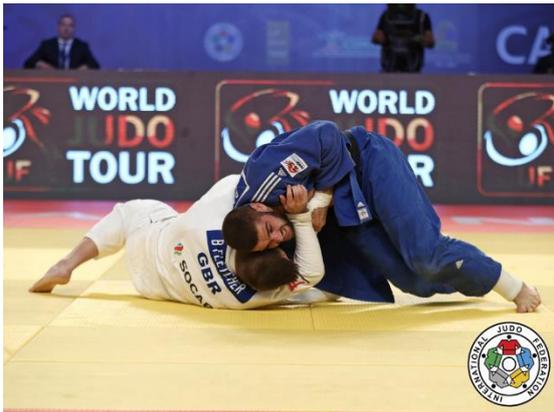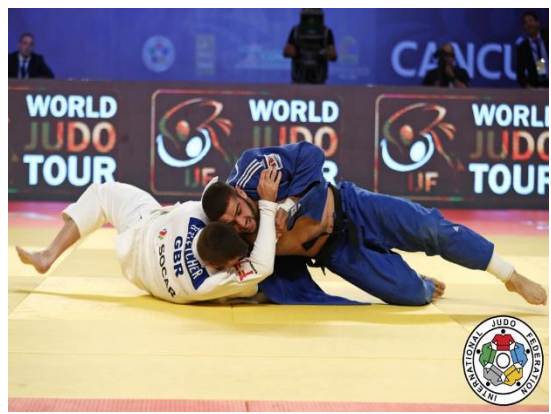
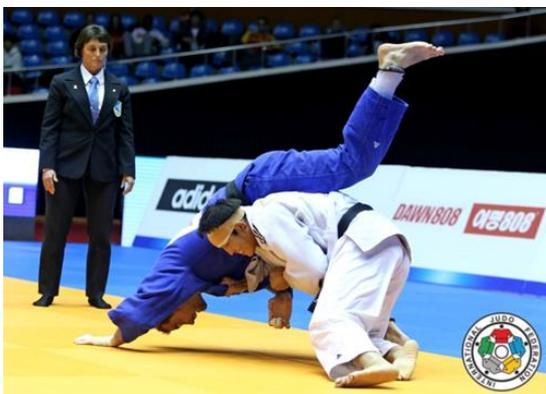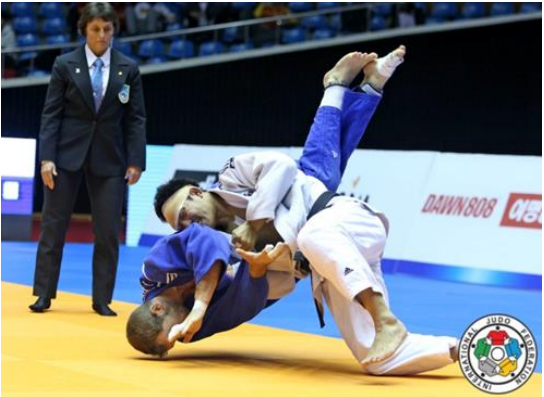
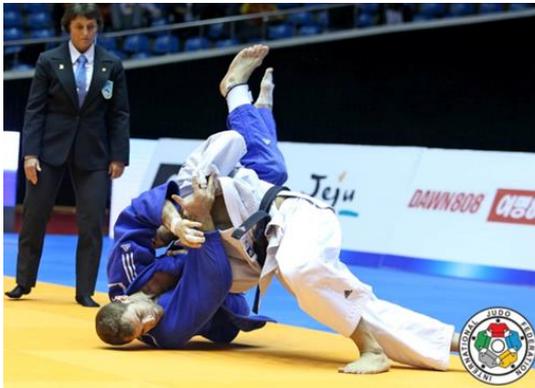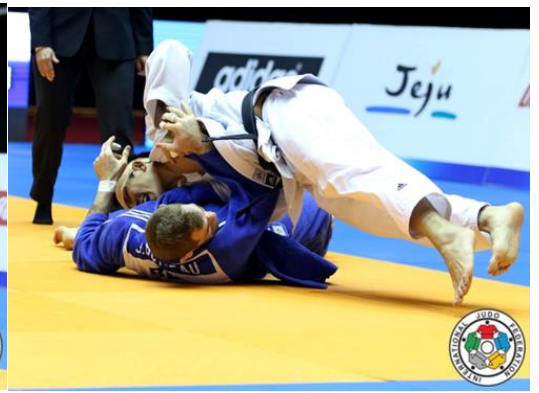

*Fig24- different Complementary Tactical Tools applied to perfect the Suwari Seoi final action*

## 8. Biomechanics and safety of complementary movements

In the fully flexed position the extensor muscle pulls at less than 7º relative to the tibia. This gives the extensor muscle a mechanical handicap relative to the ground of between 1/150 and 1/260 (it is hard to measure accurately). and the muscle has to produce a very great deal of force in order to produce a quite modest thrust on the ground.

For example, if the at the start of jump in kneeling position with feet pointed, each leg pushes on the ground with a force of about 0.05 N, it means that the muscle must be producing a force of 10-15 N more or less 100 – 200 times the starting push.

If we think to the resistance that must be overcame in a Suwari Seoi techniques: mainly the sum of Tori and Uke body's masses, for example respectively 63 and 65 Kg, then muscles needed to apply a force to move 1255,6 N   this means that each leg must push the ground with a force of 627,8  N and produces a force of 62780 N if static, dynamic situation could be more favorable but the force production must be also very high.

Even more expensive could result the same action starting from a seiza position. For these reasons more often, a different basic variation is applied in real competition, it is a form intermediate between seiza and deep kneeling, normally Tori takes support not on both legs, but differentiated support, more on one leg than on the other. This different weight distribution varies the stability of Tori which becomes more mobile and can apply complementary tools more easily. Obviously, the price of this greater freedom of movement is the potential danger due to the choice of one only support base for the dropping down action, rather than with two, in which the pressure due to the impact of the fall on the knees can decrease.

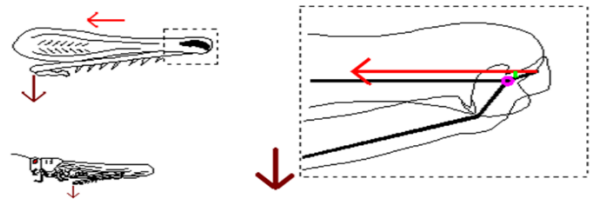

*Fig 32  quadriceps action pushing on the mat*

Deep Japanese studies have been performed to study the force production to the knee in seiza and deep kneeling, both descending and rising [52, 53,54]

The complex mathematical model is based on the evaluation of moments and external and internal forces on the hip, knee and ankle. But the solution of the system is undetermined because there are three equations with six variables (the six muscles forces) that are unknown.

However with some assumptions and simplifications it possible to decrease the number of unknown to three, and to have the force acting on the knee with good approximation.

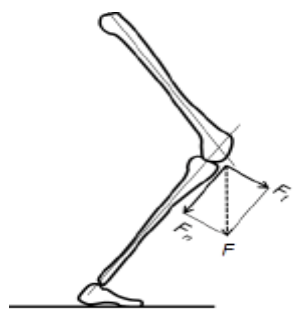

$F_n$: the normal component of knee joint force, which is parallel to the tibial axis
$F_t$: the tangential component of knee joint force, which is orthogonal to the tibial axis
$F$: the net knee joint force

*Fig.33 Forces on knee*

$$F = \sqrt{F_n^{\,2} + F_t^{\,2}}$$
$$F_n = Q' + G + H\cos\theta \qquad \{26\}$$
$$F_t = H\sin\theta$$

$$\Rightarrow F = \sqrt{Q'^2 + G^2 + H^2 + 2Q'G + 2Q'H\cos\theta + 2GH\cos\theta} \qquad \{27\}$$

The experimental results for the Knee joint force evaluated in function of the angle of knee flection are presented in the next Fig

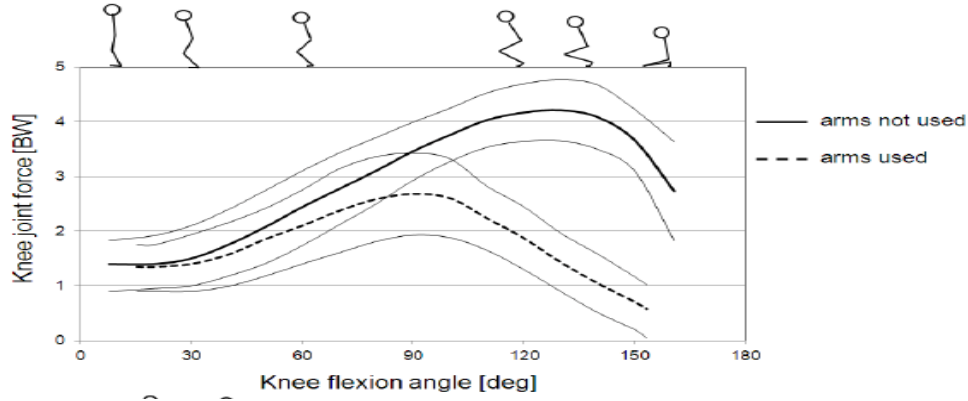

*Diag. 4 Knee joint force in rising without and with help of arms [52]*

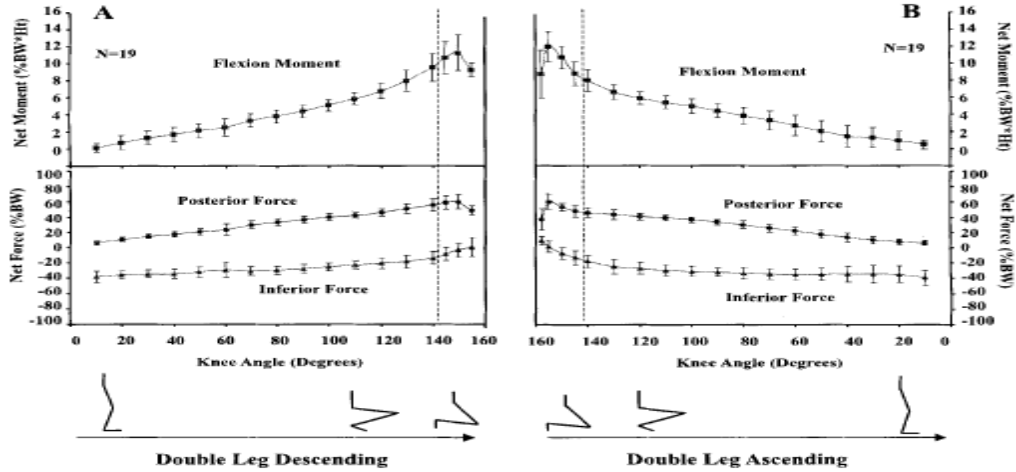

*Diag.5. Knee Joint force and momentum (ascending and descending) [53]*

The follow up to refine the throwing action of one of Suwari Seoi family techniques, the complementary movements, called Tactical Tools, can be modeled as a linear combination of different contribution lift up plus elastic force plus torsional force:

$$F_{TT} = (m_1 + m_2)a\frac{t_l}{t_{TT}} - k(L\sin\varphi)\frac{t_f}{t_{TT}} + \tau_{max}\frac{d^3}{5L}\frac{t_r}{t_{TT}} \qquad \{28\}$$

With variable in time contributions, where: $m_1$ $m_2$ = body masses of Tori and Uke, k= elastic constant of Tori thigh muscle, $\tau_{max}$ = Torque produced in the leg, d= Tori thigh muscle diameter L = Tori thigh muscle length, $t_{TT}$= time of Tactical Tool application, $t_i$= time: lift, forward, rotation.

This equation let us to be able to evaluate the order of magnitude of the effort in the knee, in order to verify if the ligament structures of Tori's knees can be possibly put at risk during the execution of these movements.

If we put inside the right physiological values, for a theoretical movement performed in 0,26 s divided into three sub-movements of the same time duration, all this applied to two athletes of the category up to 65 kg, a category that often applies these techniques.

the critical nature of the complementary movements, on the safety of the knees, is shown through a purely theoretical calculation in order to focus the attention of the teachers on a correct study of this part of the technique. Remembering the previous equation {28} and that the physiological parameters are related to the most used thigh extensor muscles (quadriceps femoris for women, and hamstring group for male) [55]

$$F_{TT} = (m_1 + m_2) a \frac{t_l}{t_{TT}} - k(L \sin \varphi) \frac{t_f}{t_{TT}} + \tau_{max} \frac{d^3}{5L} \frac{t_r}{t_{TT}}$$

The theoretical evaluation, for a 65 kg against a 62 kg, shows us that for Female and Male the relative results are: 1750 N and 1560 N, with rotational contribution about 400 N. Remembering that normal resistance to traction is: for ACL 2000 N, PCL 1500 N and lateral structures around 500 N [56] we, easily, understand how delicate the perfect execution of this part is.

The last notation is that, during the competitive application, most of rotatory effort applied, in safe way, are performed by the trunk-hip complex, lightening the task of the knees.

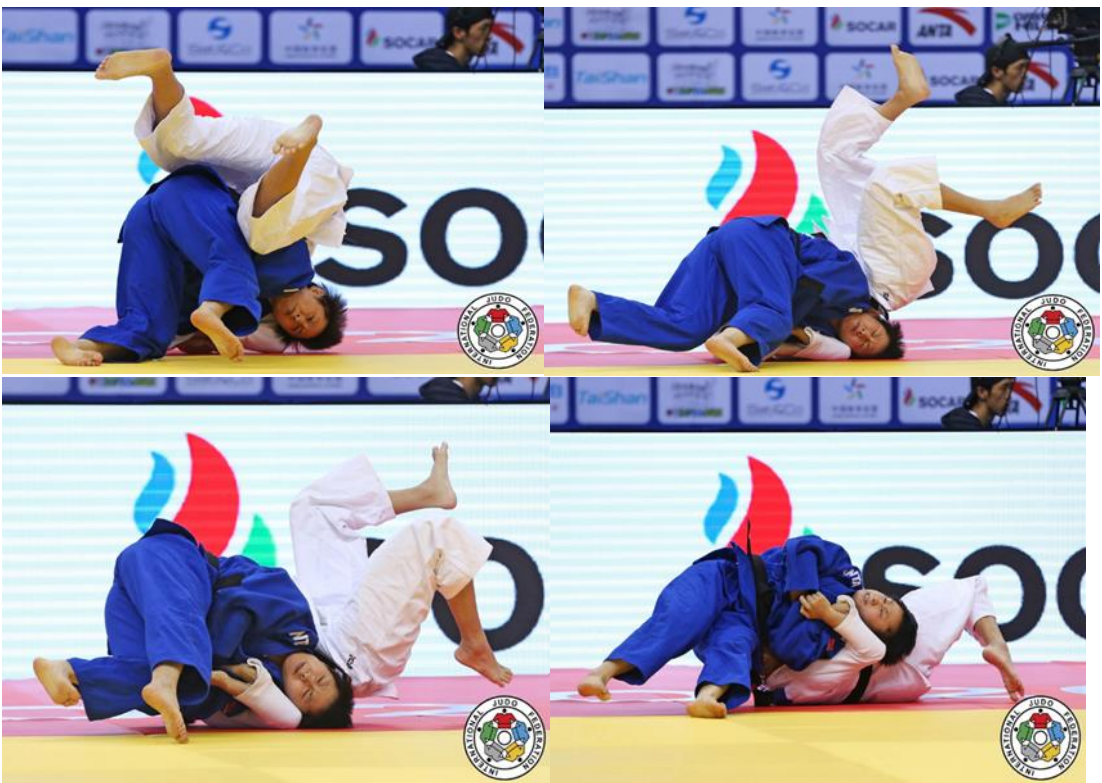

*Fig 34-37 Torsional application in complementary movements*

## 9. The experimental protocol

Each subject performed two specific trials with three subcases, and six performances of throw. All these trials, for four times, for obvious statistic reasons, to decrease the variance among performances.

This means that each child Tori performed 24 actions, the national athletes, for time and training reasons, only two trials, that is 12 actions. But both for safety reasons and time, Children performed the trials in three days, Athletes in two days.

Each trial was divided in three special performances: one to perform suwari seoi alone, two with uke without throwing him, and three with uke, throwing him to complete the technique.

These diversification is made to study the influence of the ukè body, on the speed of falling of Tori's knees in the dynamic equilibrium of the couple.

Thus Tori falls first on his own, then with the holds to uke's judogi and finally by performing the projection of uke's body.

The first trial is with falling down with both feet pointed touching the tatami, then the mechanics of this way of application it is a fall with control, specifically a mechanic paradox ( see biomechanics paragraph 6 ), and the area of percussion is smaller than the other way to touch the Tatami, only the tips of pointed feet and small tibial area, we take time of fall's trajectory, to evaluate the mean speed of fall.

The second trial is the other way to apply Suwari Seoi in competition, jumping with the feet stretched out not touching at all the tatami , in that case the mechanics of the action is a free fall and the contact area is larger, long tibial surface projection and back of the feet

The impact surface measurements are based on the dynamic heat conduction between human body and surface layer of the tatami and followed by the natural cooling for convection.

Few studies using infrared thermography have been devoted to sports performance diagnostic and to sports pathology diagnostic, no study are developed for sport safety and prevention.

It is well known that sports activity induces a complex thermoregulation process where part of heat is given off by the skin of athletes. As not all the heat produced can be entirely given off, there follows a muscular heating resulting in an increase in the skin superficial temperature. In particular, the IRT method will enable, in the long term, to quantify the heat loss.

This research is focalized on the capture of the thermal image of surface body contact left by Tori after the fall to throw by Suwari Seoi and on its geometrical measure, in both cases feet pointed and feet stretched.

When Tori body falling touches the tatami it leaves one thermal track produced by dynamic thermal conduction.

This not visible, thermal track disappear very fast due to the cooling by convention of the surface layer, when the body leaves the tatami.

In formula the eating of Tatami surface due to conduction between human body contact and Tatami surface is driven by the classical equation:

$Q_{cond} = -k\nabla T$  {29}

The cooling [49] of Tatami surface is driven by the natural convection in closed environment

$Q_{conv} = \rho c_p V T$  {30}

Remembering also that the general equation for convection is the well-known:

$Q_{conv} = h(T - T_0)$  {31}

It is possible to obtain the differential equation that shows the variation in time of temperature during cooling.

$$\frac{dT}{dt} = \left(\frac{h}{\rho c_p \Delta x}\right)(T_0 - T) \quad \{32\}$$ The solution of this differential equation is:

$$T = T_0 + (T_1 - T_0)e^{-\left(\frac{h}{\rho c_p \Delta x}\right)t} \quad \{33\}$$

The evaluation of the inverse of the exponent gives us the order of magnitude of the time of cooling phenomenon that is, replacing the parameters of the materials, about 45 s.
That is very slow.[57, 58]
The research idea is to capture by a fast and sensible thermal camera this evanishing image of the contact surface, measure it and evaluate from the safety point of view, the stress received by body that is the maximum impact force divided the measure of evanishing thermal image of contact surface. [ 59, 60]
Then the protocol is very simple on a prepared Tatami see next figure both children and or athletes perform the two trials of throws in the two-ending condition.
Feet pointed, feet stretched.
The lay out of the research this time is organized with two tatamis prepared and a cooling system to cool the tatami and speed-up the research time.

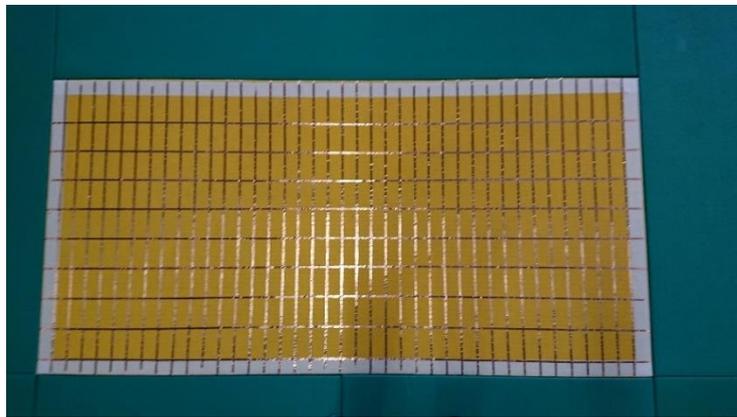

*Fig.38 Tatami prepared with a special chopper tape useful for thermal images*

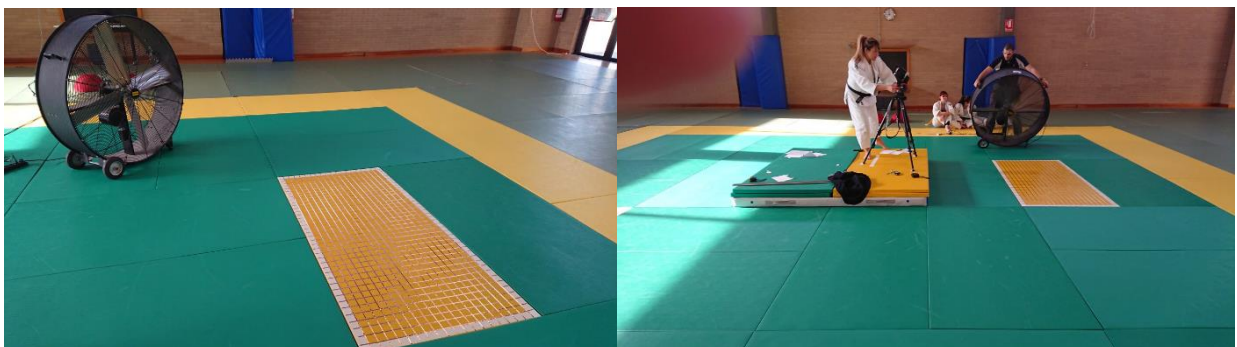

*Fig.39-40 Cooling system and tatami*

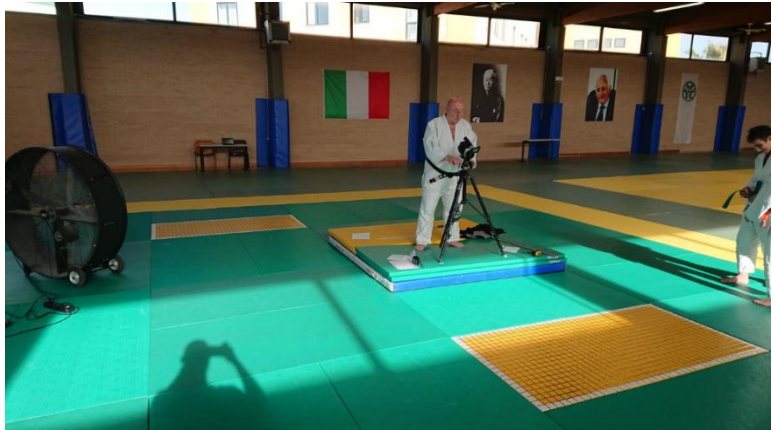
*Fig.41 Complete lay-out of Research with two tatami prepared*

Two digital chronometers are utilized to evaluate the flight time or the trajectory time of the subjects, to have a better evaluation.
Then with the thermal camera are sampled the surface of contact for each trial.
In the next figures there are shown some thermal images of the contact surface relative to the two stiles analyzed, it is easy to see the difference in surface obtained.
 by the Japanese Thermal Camera AVIO 600 from Nippon Avionics.
Equipped with the software InfReC Analyzer 9500.

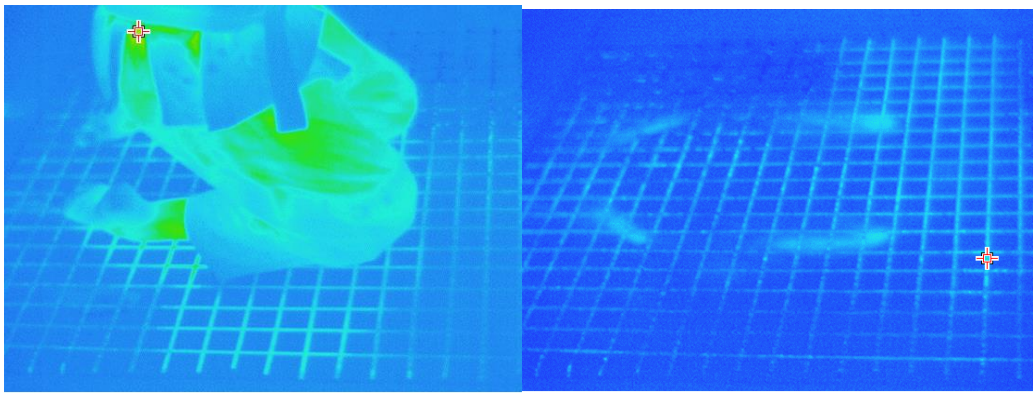
*Fig42-43 Children Feet Stretched and Thermal image*

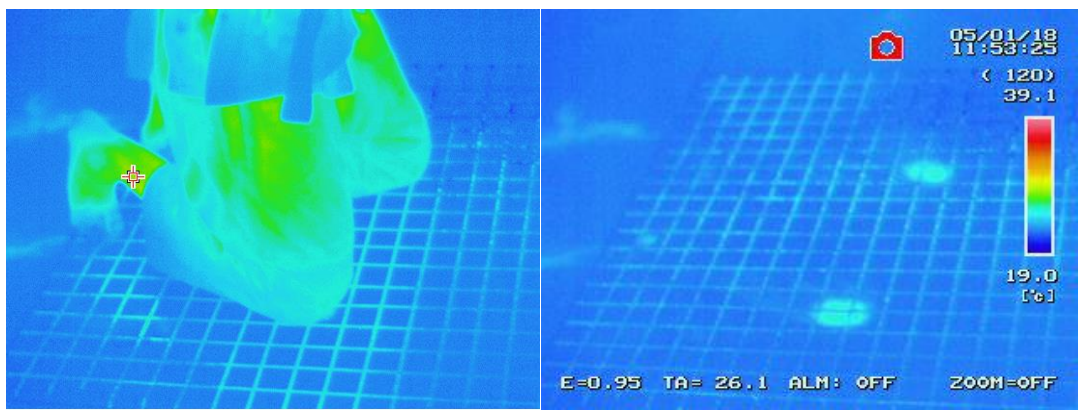
*Fig44-45 Children Feet Pointed and Thermal Image*

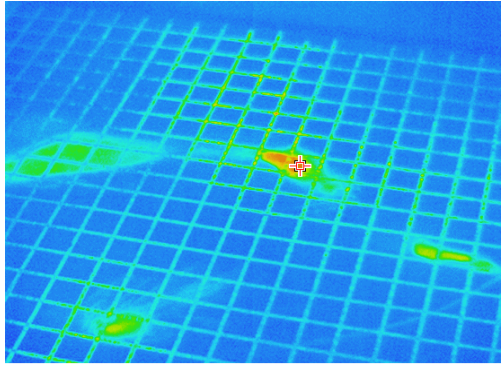 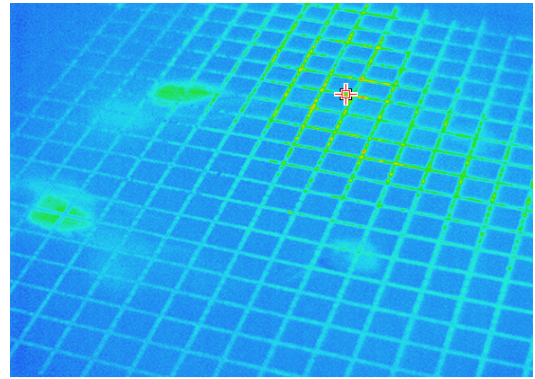

*Fig.46 Adult Feet stretched Preponderance from one side*     *Fig47  Adult knees feet pointed*

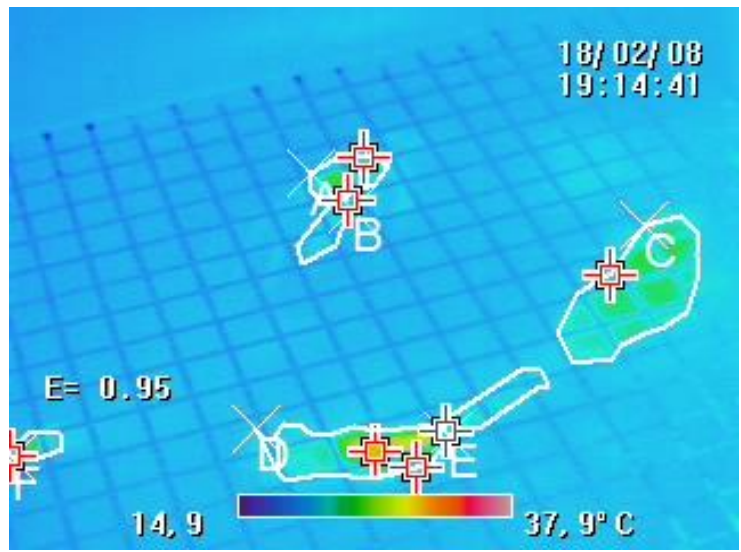

*Fig48  measurement of surface area of Suwari with right preponderance*

| YEAR | H BODY | W | H Knee-heel | Costeff | DUBOIS |
|---|---|---|---|---|---|
| | m | Kg | cm | m² | m² |
| male | | | | | |
| 1999 | 1,56 | 55,3 | 44 | 1,25643 | 1,53829 |
| 2001 | 1,77 | 71 | 47 | 1,44596 | 1,87469 |
| 2000 | 1,70 | 56,5 | 49 | 1,27235 | 1,65219 |
| 2002 | 1,70 | 54 | 50 | 1,23888 | 1,62072 |
| 2006 | 1,50 | 45 | 42 | 1,08148 | 1,36977 |
| Female | | | | | |
| 2005 | 1,65 | 50 | 48 | 1,18285 | 1,53498 |
| 2000 | 1,60 | 58,7 | 45 | 1,30087 | 1,60703 |
| 2003 | 1,65 | 55 | 43 | 1,25241 | 1,59843 |

*Tab11 Children Data*

| YEAR | H BODY | W | H Knee-heel | W+ J | DUBOIS | Dubois J |
|---|---|---|---|---|---|---|
| | m | kg | cm | kg | m² | m² |
| male | | | | | | |
| 1999 | 1,67 | 63 | 49 | 64,5 | 1,70827 | 1,72543 |
| 1998 | 1,69 | 63,8 | 51 | 66,5 | 1,73233 | 1,76312 |
| 1993 | 1,72 | 63,5 | 50 | 66,1 | 1,75107 | 1,78118 |
| 1993 | 1,69 | 71,6 | 48 | 74 | 1,81937 | 1,84505 |
| 1989 | 1,90 | 103,5 | 57 | 106,3 | 2,31639 | 2,34282 |
| 1991 | 1,87 | 104 | 58 | 106,7 | 2,29451 | 2,31964 |
| 1999 | 1,73 | 75 | 50 | 77 | 1,88734 | 1,90857 |
| 1998 | 1,7 | 70,8 | 50 | 73,2 | 1,81846 | 1,84442 |
| 1996 | 1,82 | 93 | 54 | 96,4 | 2,14547 | 2,17847 |
| 1990 | 1,9 | 110 | 55 | 115 | 2,37714 | 2,42248 |
| 1997 | 1,73 | 71,5 | 46 | 73 | 1,84939 | 1,86579 |
| 1998 | 1,76 | 75 | 51 | 78,2 | 1,91101 | 1,94525 |
| Female | | | | | | |
| 1998 | 1,61 | 50,7 | 46 | 53 | 1,51685 | 1,54572 |
| 1999 | 1,6 | 50,1 | 47 | 52,8 | 1,50239 | 1,53629 |
| 1997 | 1,63 | 54 | 47 | 56,7 | 1,57206 | 1,60500 |
| 1998 | 1,49 | 48,2 | 47 | 50,3 | 1,40352 | 1,42919 |
| 1996 | 1,64 | 52 | 47 | 53,2 | 1,55392 | 1,56906 |
| 1995 | 1,58 | 59,1 | 45 | 61,9 | 1,59704 | 1,62876 |

*Tab 12 National Team data*

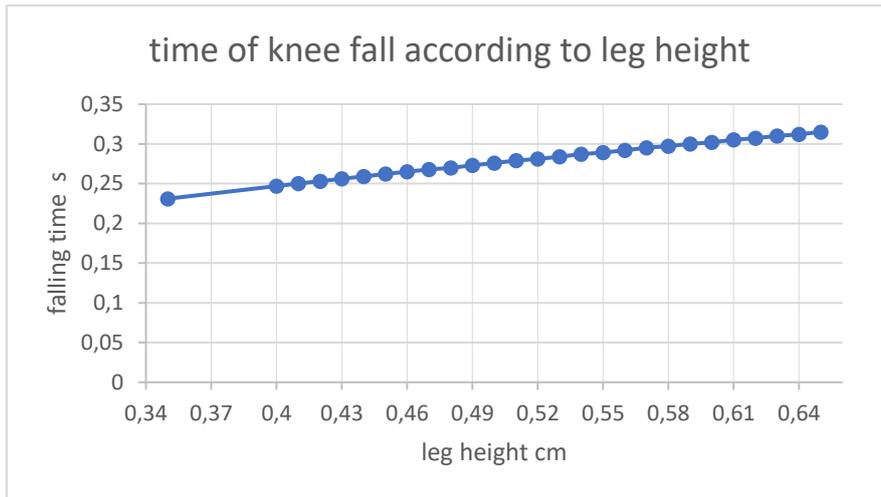

*Diag.6 Feet Pointed falling time variation according to leg height*

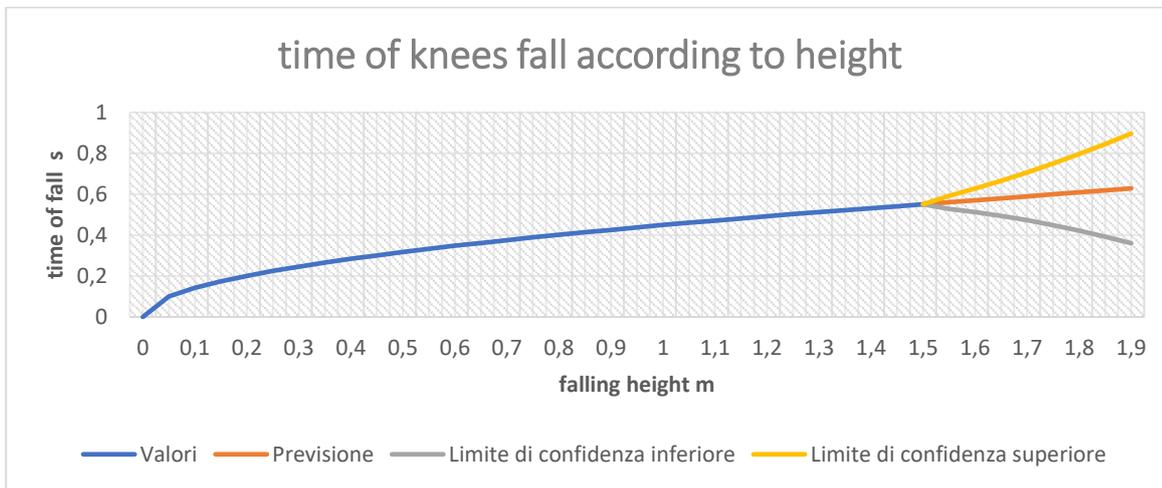

*Diag. 7 free fall time of fall according to falling height*

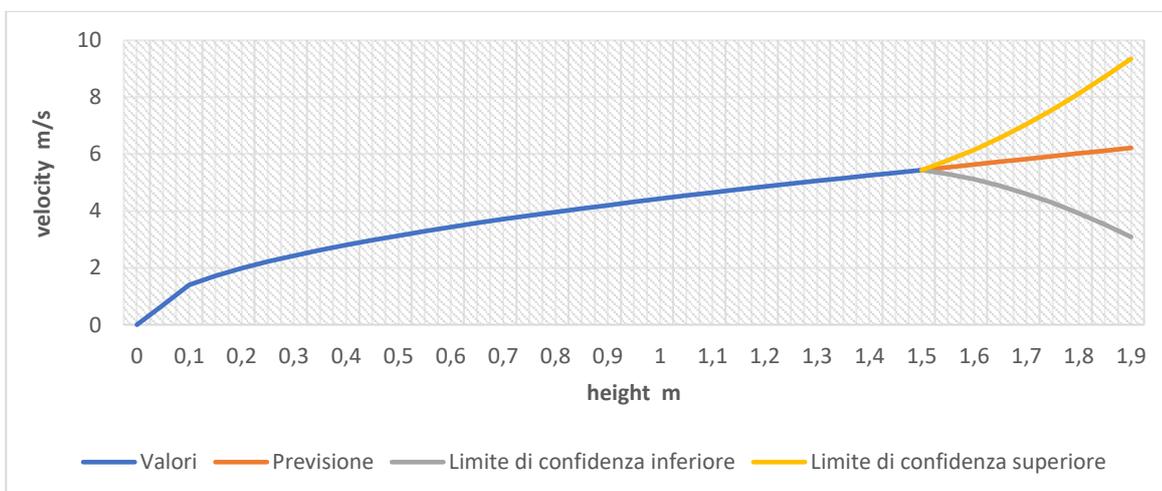

*Diag 8 free fall knees velocity according to height*

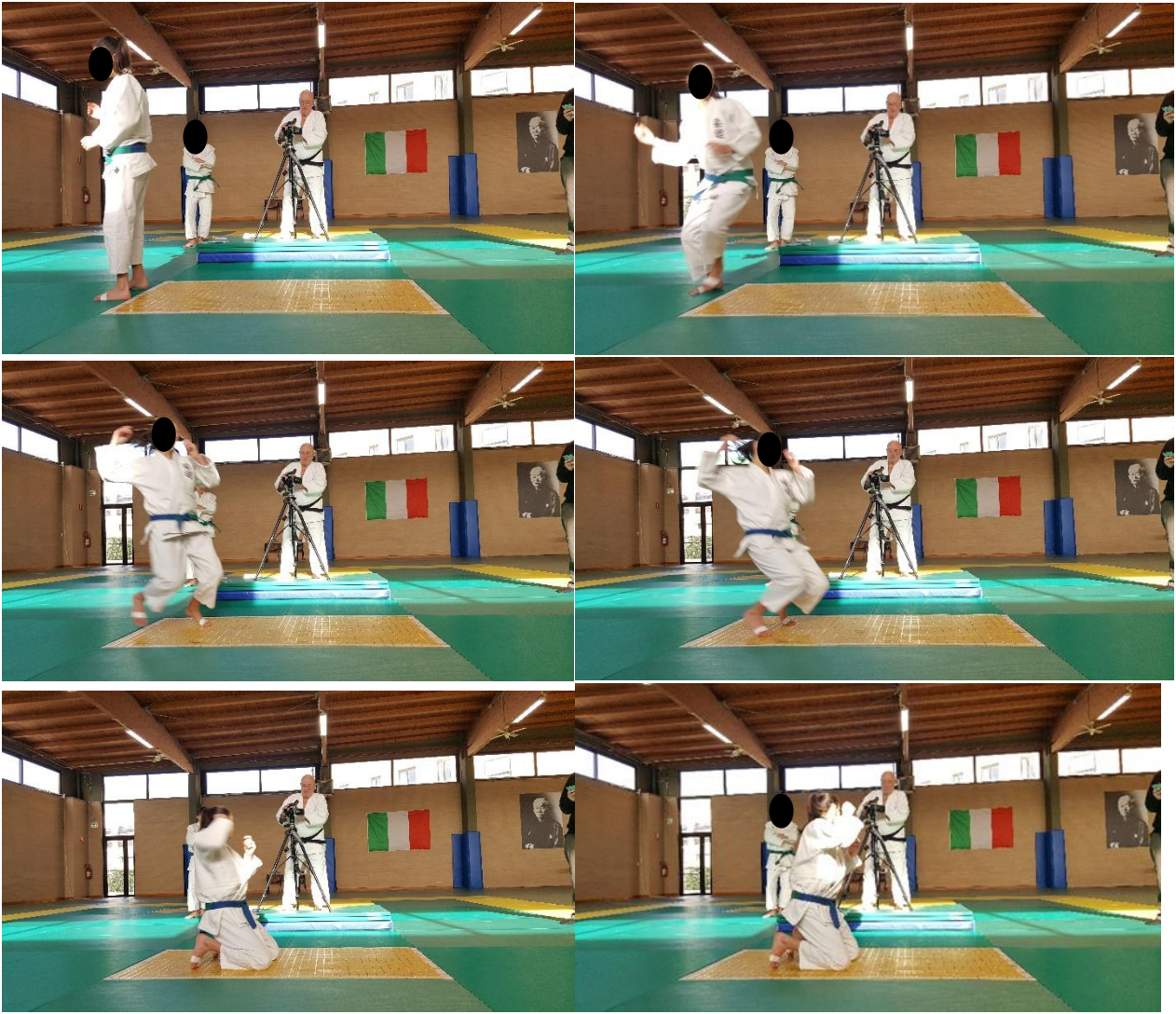

*Fig49-58 Experimental Protocols : Suwari Seoi alone and with no throwing action*

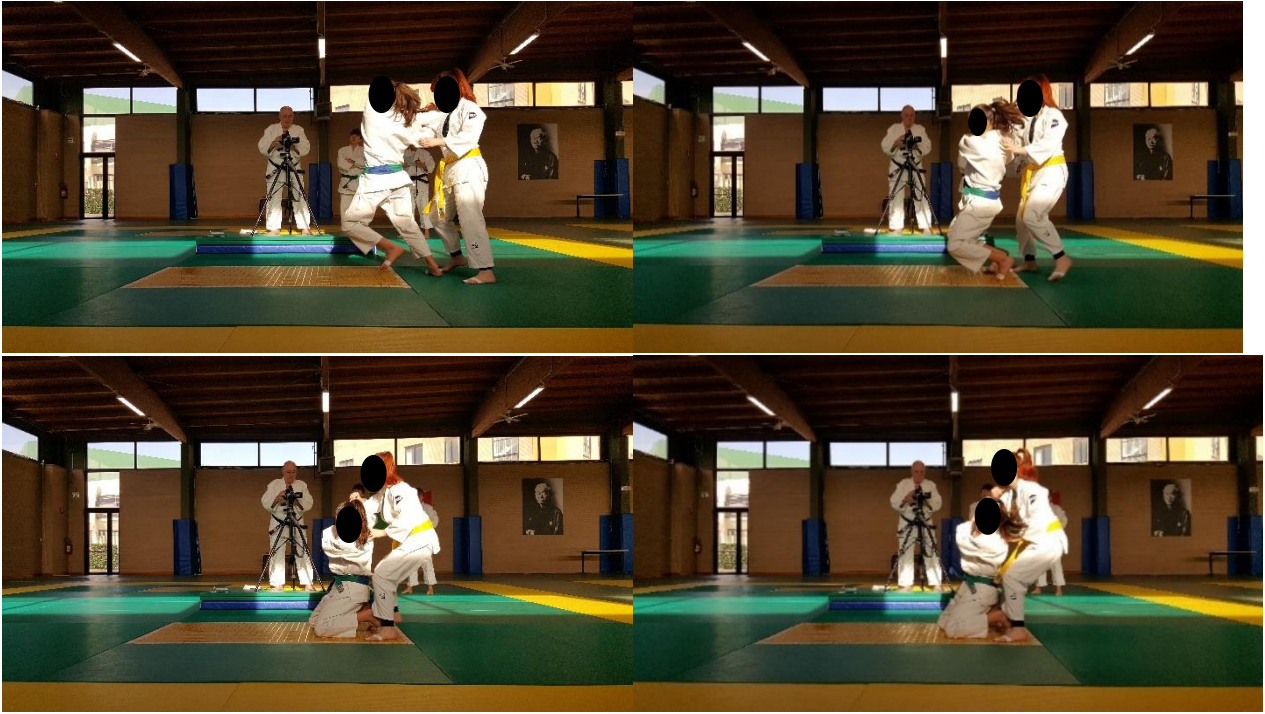

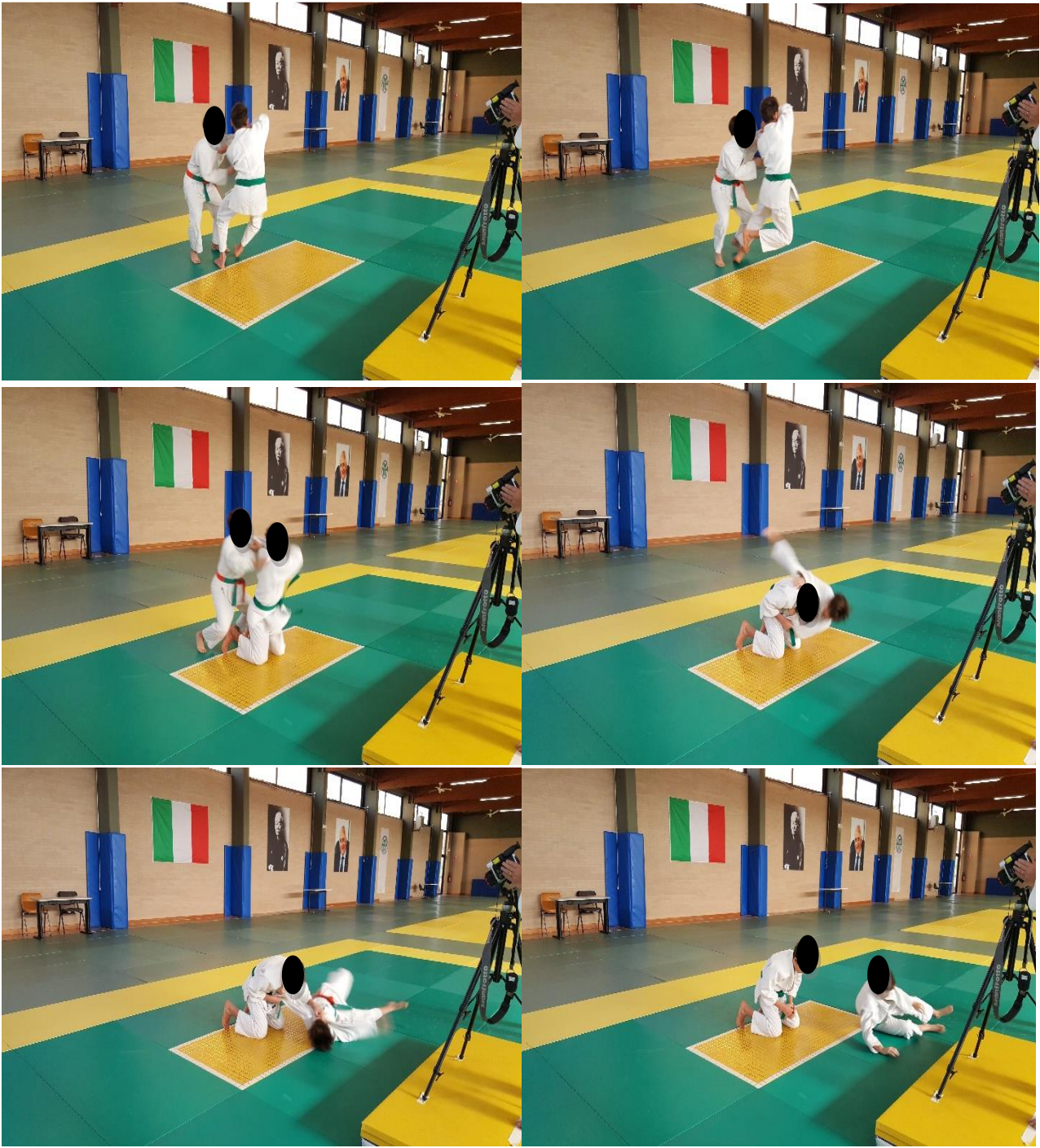

*Fig. 59-64 Experimental Protocol : Suwari Seoi with throwing action*

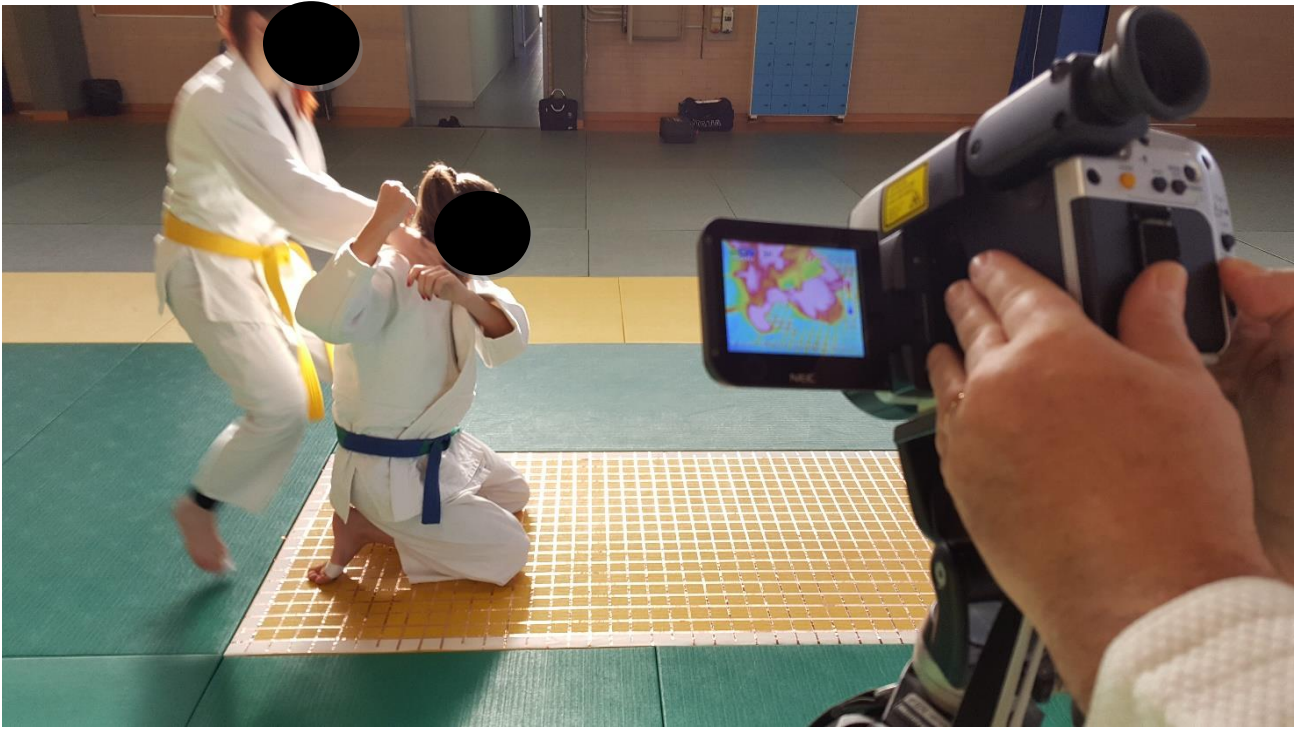

*Fig 65-66 Thermal capture of athlete*

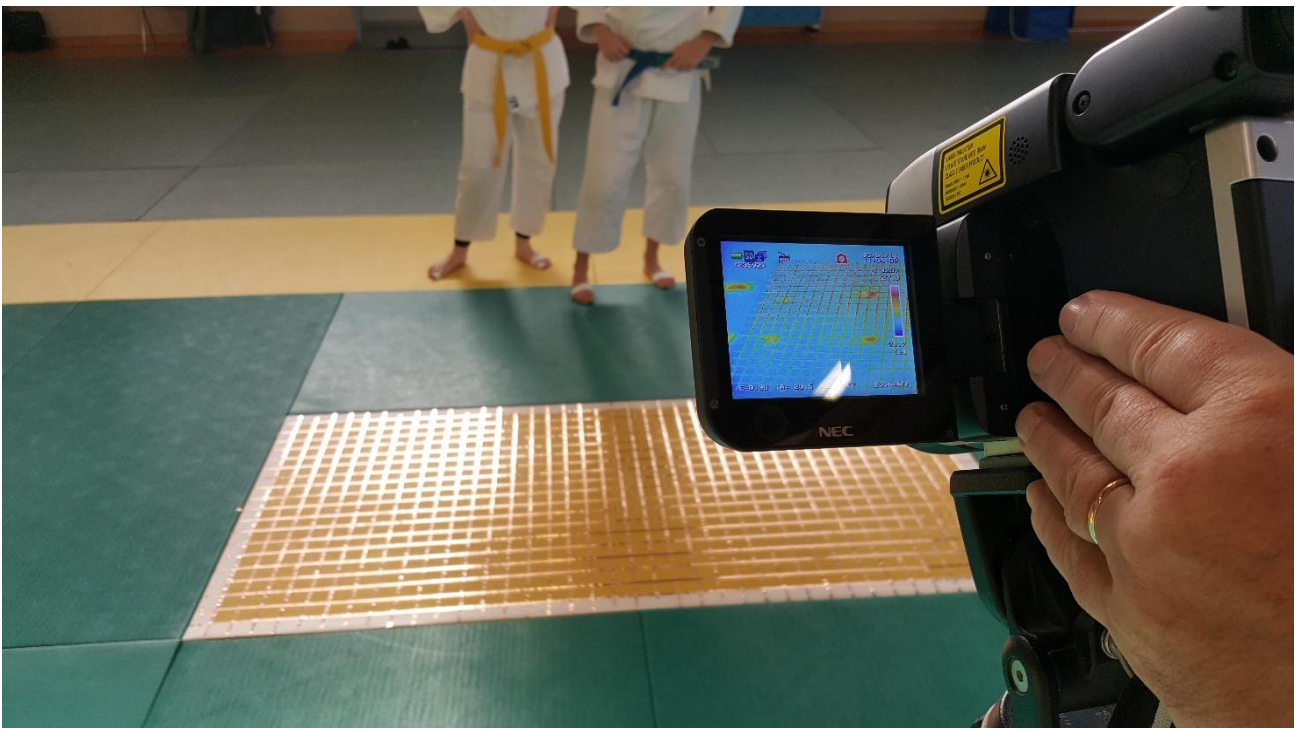

*Fig 67-68 thermal capture of contact knees image*

*10. The experimental results*

The in-depth work carried out on the techniques: analysis of 437 thermal films, (223 of the children and 214 of the national Athletes) each of which averaged about 60 frames allowed to evaluate the average of the contact surfaces between the four tests performed for the children and the two performed for adults both in the case of execution with pointed feet and in the case with stretched feet.

The average contact surfaces obtained have been divided for the overall body surfaces, thus obtaining the surface contact percentage for each style. These percentages were compared between children and adults, in order to obtain reference information on the greater or lesser danger of the style of the technique as a function of the age of Tori.

The average time of fall or flight, (two measures) for each style made it possible to calculate the experimental fall velocity of Tori's knees.

This experimental speed was compared to the theoretical one, measured by the equations of the biomechanical model.

The biomechanical model made it possible, also to calculate the impact force in the worst conditions, (safety analysis), which resulted in the worst pressure, which weighs on Tori's knees for each style, and the ground reaction force of the mat that is received, assessing its dangerousness.

All measures are compared among children and adult to single out any difference.

This comparison in term of safety is based on the SBA ( Surface Body Area) calculation.
 Normally for children it is utilized the Costeff formula [99] more accurate than the, well known, formula of Du Bois and Du Bois [pp] that overestimate the body surface for children. However we use the Du Bois and Du Bois formula applied for children, only in the case of the comparison with adults, because in this way, the measures of the overestimate surface increases (for safety)  the pressure on the children knees, and give us amore safe way to compare.

Some interesting safety evaluations are singled out from the specific experimental protocol.

At first children show a constant decrement of the falling velocity in both two approaches: feet pointed and feet stretched, starting from the exercise alone, ending with the projection exercise. This behavior is very important from the safety point of view.

It is easily explained, both :  by the poor technical ability and by the dynamical equilibrium between Tori and Uke, in other words the presence of the Uke Body slows down the knee speed of Tori who is relented by the action of moving the body mass of Uke.

In this way the final pressure will be less dangerous for the children knees.

In the following tables and diagrams there are shown two examples of this decreasing in speed averaged on five trials.

Equivalent results were found for the 95% of the other children the last 5% was almost equal as knees falling speed .

| Average knees' speed in the trials of the two styles of suwari seoi, same boy M     m/s | | |
|---|---|---|
| *Tori alone* | *With Uke no throwing* | *Throwing Uke* |
| 1,88 | 1,38 | 1,23 |
| 1,46 | 1,56 | 1,34 |
| 2,13 | 1,56 | 1,42 |
| 1,62 | 1,95 | 1,17 |
| 2,04 | 1,04 | 1,27 |
| 1,95 | 1,51 | 1,23 |
| 2,23 | 1,42 | 1,51 |
| 1,88 | 1,62 | 1,17 |
| 1,77 | 1,61 | 1,29 |
| 2.02 | 1,39 | 1,29 |

*Tab13  average knees' falling speed (white Feet Pointed, Blue Feet stretched) Male*

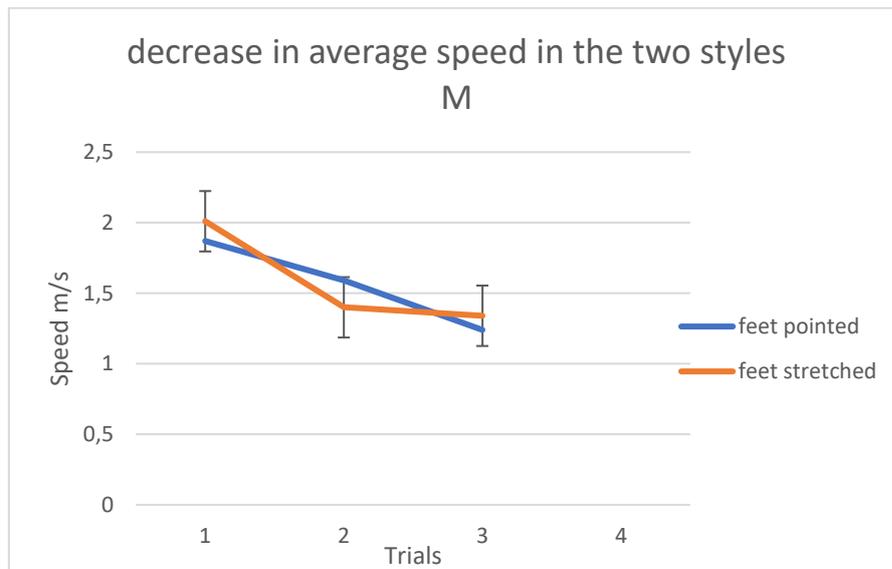

*Diag. 9 Example of Children average speed decreasing M*

| *Average knees' speed in the trials of the two styles of suwari seoi, same girl F m/s* | | |
|---|---|---|
| *Tori alone* | *With Uke no throwing* | *Throwing Uke* |
| *1,28* | *1,44* | *1,48* |
| *1,58* | *1,4* | *1,16* |
| *1,75* | *1,32* | *1,25* |
| *1,96* | *1,44* | *1,4* |
| *1,48* | *1,48* | *1,36* |
| *1,36* | *1,36* | *0,94* |
| *1,63* | *1,48* | *1,13* |
| *1,88* | *1,48* | *1,4* |
| *1,64* | *1,4* | *1,32* |
| *1,58* | *1,39* | *1,29* |

*Tab 14 average knees' falling speed (white Feet Pointed, Pink Feet stretched) Female*

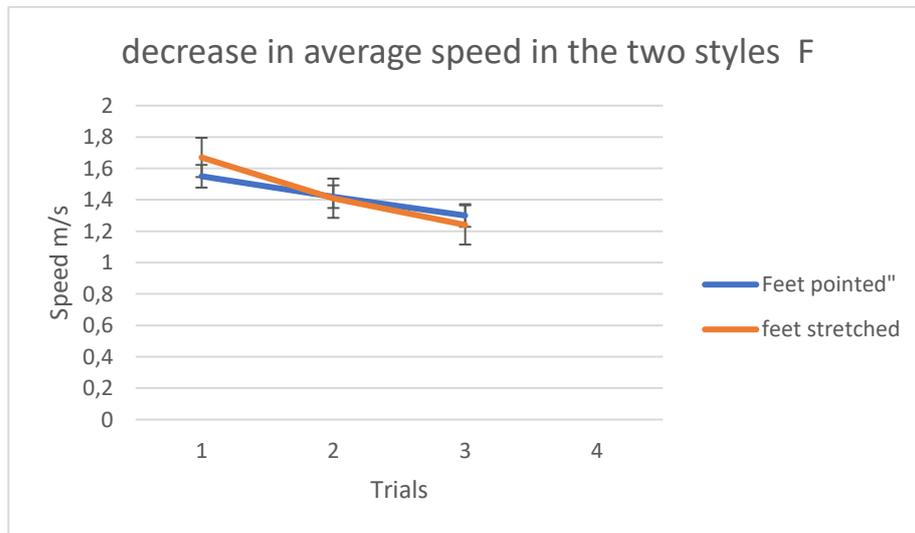

*Diag,10 Example of Children average speed decreasing F*

More variable trend is found in the adults of the Italian National Team, that possess a greater technical ability and more accentuated automatisms.
This means that the skill ability is, at high level, the discriminating factor between athlete and athlete.
Despite the small numerical sample, we give, in indicative way, the percentage trends found between the athletes of the national team divided as males and females.
In fact, among male about 63% has showed the expected decrease, while 7% more or less constant speed, and a 30% increase in speed in correspondence with the acquired competitive technique. Among female 61% showed decrease, 30% showed increase and 9% more or less constant speed.

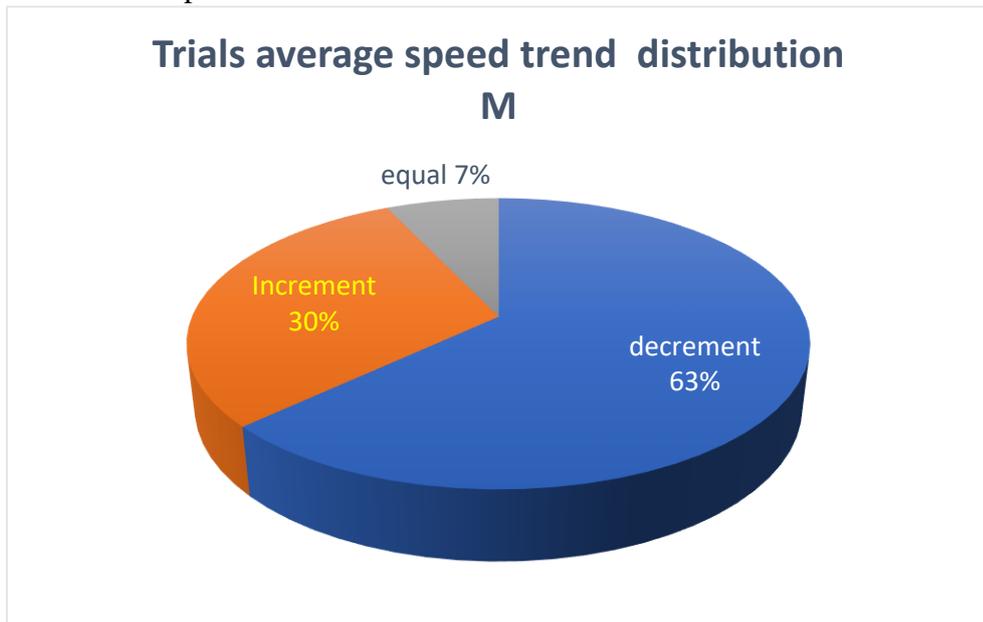

*Fig69 Indicative average speed trend, National Team (Male)*

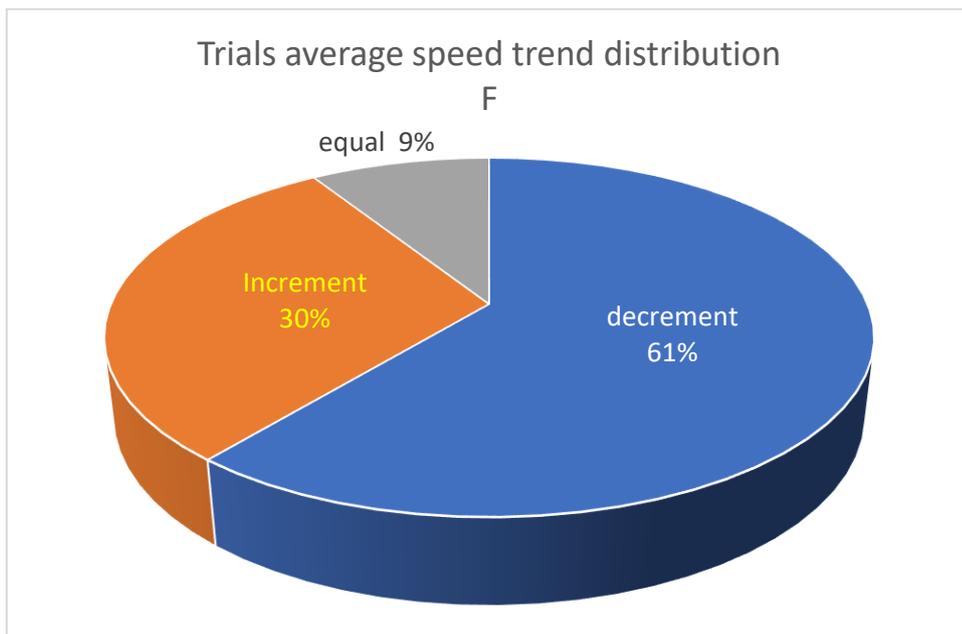

*Fig 70 Indicative average speed trend, National Team (Female)*

Connection between speed and skill of the athletes was also able to highlight, in which the fastest ever was a former world championship winner.

| average speed in the trials of the two styles of suwari seoi, same Athlete M m/s |||
| --- | --- | --- |
| Tori alone | With Uke no throwing | Throwing Uke |
| 1,9 | 2,31 | 2,31 |
| 2,5 | 1,3 | 2,04 |
| 1,13 | 2,12 | 2,37 |
| 1,9 | 2,31 | 2,31 |

*Tab 15 average speed increasing M*

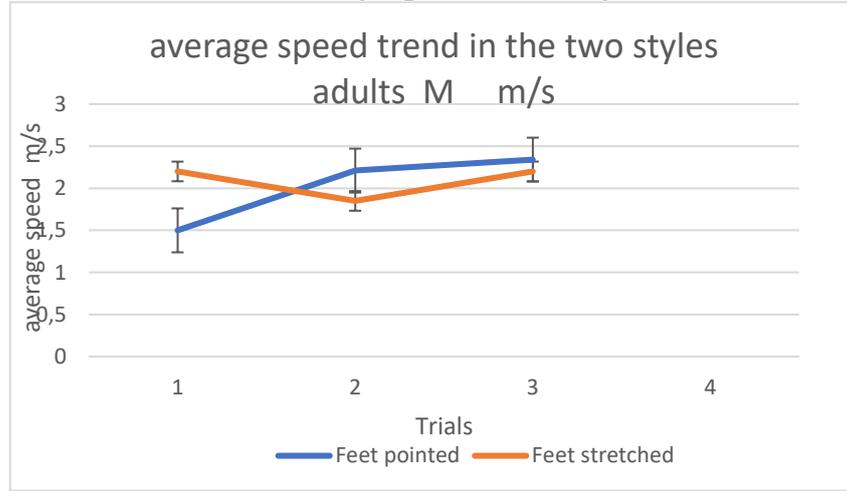

*Diag. 11 speed trend*

| average speed in the trials of the two styles of suwari seoi, same girl F m/s |||
| --- | --- | --- |
| Tori alone | With Uke no throwing | Throwing Uke |
| 1,64 | 2,13 | 1,95 |
| 1,88 | 2 | 2,24 |
| 1,62 | 1,84 | 1,95 |
| 1,88 | 2,23 | 2,35 |

*Fig.16 average speed increasing F*

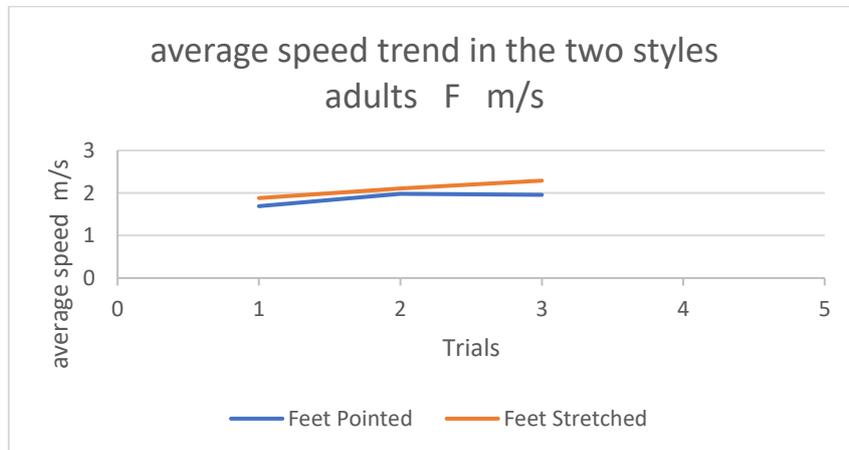

*Diag.12 speed trend*

The safety aspect of the throw for children and even for athletes, first of all it is guaranteed by the experimental data that for the majority of both boys and athletes, the average fall speed of the knees decreases when uke is projected. The next step will be to measure the average standard judoka child [63], the stress that are produced at the knees in the two styles at impact. [64]. At the end we evaluate the AIS ( *Abbreviated Injury Scale*) for the determined compression, to see if there is danger in the application of Suwari Seoi Family.

$$AIS = -3{,}78 + 19{,}56 C \qquad C = \Psi/E = F/AE$$

The Compression factor is evaluated by the stress received divided by the Young Modulus of Tatami  E= 0,44  GPa

In safety term considering the maximum available values of forces in the two cases feet pointed and feet stretched, that are Respectively 3 BW and  7BW . We can evaluate the results at light of AIS,

|  | CHILDREN | | ADULTS | |
|---|---|---|---|---|
|  | MALES | FEMALES | MALES | FEMALES |
| MEAN S FEET POINTED | 0,014 m² | 0,012 m² | 0,0167 m² | 0,0132 m² |
| MEAN S FEET STRETCHED | 0,022 m² | 0,017 m² | 0,047 m² | 0,029 m² |
| BSA MEAN DUBOIS | 1,5787 m² | 1,614 m² | 1,9456 m² | 1,5242 m² |
| category Es. A | 1,36977 | | | |
| category Es. B | 1,53498 | | | |
| cadets | 1,87469 | 1,59843 | | |
| juniors | 1,53829 | 1,62961 | 1,811481 | 1,47425 |
| seniors | | | 2,07980 | 1,57434 |
| MEAN WEIGHT | 54,45 Kg | 55,5 Kg | 78,8 Kg | 52,35 Kg |
| category Es. A | 45 | | | |
| category Es. B | 54 | | | |
| cadets | 63,5 | 52,5 | | |
| juniors | 55,3 | 58,6 | 69,52 | 49,6 |
| seniors | | | 88,08 | 55 |
| MEAN WEIGHT WITH JUDOJI | 55,7 Kg | 56,5 Kg | 81,3 Kg | 54,45 Kg |
| category Es. A | 46 | | | |
| category Es. B | 56,1 | | | |
| cadets | 64,6 | 53,4 | | |
| juniors | 56,4 | 58,6 | 71,8 | 52 |
| seniors | | | 90.8 | 56,9 |
| MEAN YEARS all subjects | 2001 | 2001 | 1995 | 1997 |

*Tab17  Comparison between children and Adult mean data*

| children | Male f.p.s (m²) | Male f.s.s (m²) | Max force Fp 3BW (N) | Max force Fs 7BW (N) | Mass Kg | Stress Fp MPa | Stress Fs MPa | Compression Adimensional 10⁻³ | |
|---|---|---|---|---|---|---|---|---|---|
| category Es. A | 0,010 | 0,02 | 1324,35 | 3090,15 | 45 | 0,13 | 0,15 | 0,29 | 0,34 |
| category Es. B | 0,014 | 0,022 | 1471,51 | 3433,5 | 50 | 0,10 | 0,15 | 0,22 | 0,34 |
| cadets | 0,017 | 0,024 | 2089,53 | 4875,57 | 71 | 0,12 | 0,19 | 0,27 | 0,43 |
| juniors | 0,016 | 0,021 | 1656,9 | 3866,12 | 56,3 | 0.1 | 0,18 | 0,22 | 0,4 |
| *Medium standard Children* | *0,014* | *0,022* | *1618,65* | *3776,85* | *55* | *0,11* | *0,17* | *0,25* | *0,38* |
| Athletes | Male f.p.s (m²) | Male f.s.s (m²) | Max force Fp 3BW (N) | Max force Fs 7BW (N) | Mass Kg | Stress Fp MPa | Stress Fs MPa | Compression Adimensional 10⁻³ | |
| juniors | 0,0158 | 0,04 | 2113,07 | 4930,5 | 71,8 | 0,13 | 0,12 | 0,29 | 0,27 |
| seniors | 0,0174 | 0,054 | 2672,2 | 6235,2 | 90,8 | 0,15 | 0,11 | 0,34 | 0,25 |
| *Medium Standard Athletes* | *0,0167* | *0,047* | *2392,6* | *5582,8* | *81,3* | *0,14* | *0,12* | *0,31* | *0,26* |

*Tab 18 Male Children -Athletes results comparison*

fps =feet pointed surface; fss = feet stretched surface

Fp= Feet Pointed; Fs = Feet stretched

| children | Female f.p.s (m²) | Female f.s.s (m²) | Max force Fp 3BW (N) | Max force Fs 7BW (N) | Mass Kg | Stress Fp MPa | Stress Fs MPa | Compression Adimensional 10⁻³ | |
|---|---|---|---|---|---|---|---|---|---|
| cadets | 0,011 | 0,014 | 1545 | 3605,1 | 52,5 | 0,14 | 0,25 | 0,31 | 0,57 |
| juniors | 0,013 | 0,020 | 1724,5 | 4024 | 58,6 | 0.13 | 0,20 | 0,29 | 0,44 |
| *Medium standard Child* | *0,012* | *0,017* | *1633,365* | *3811,185* | *55,5* | *0,135* | *0,22* | *0,30* | *0,49* |
| Athletes | Female f.p.s (m²) | Female f.s.s (m²) | Max force Fp 3BW (N) | Max force Fs 7BW (N) | Mass Kg | Stress Fp MPa | Stress Fs MPa | Compression Adimensional 10⁻³ | |
| juniors | 0,0122 | 0,023 | 1530,3 | 3570,8 | 52 | 0,12 | 0,14 | 0,27 | 0,34 |
| seniors | 0,0142 | 0,035 | 1674,5 | 3907,3 | 56,9 | 0,15 | 0,11 | 0,26 | 0,24 |
| *Medium Standard Athlete* | *0,0132* | *0,029* | *1602,46* | *3739,081* | *54,45* | *0,12* | *0,13* | *0,27* | *0,29* |

*Tab 19 Female Children -Athletes results comparison*

fps =feet pointed surface; fss = feet stretched surface

Fp= Feet Pointed; Fs = Feet stretched

The AIS (0) values both for Children and Athletes are always negative this means: Extremely Lightweight Trauma.
Then there are nor silent or evident trauma but only light contusion.
From that we can evaluate that both application of Suwari Seoi Family: (Feet Pointed and Feet Stretched), if right applied, are safe for all Judokas, from Children in Dojo to High Level Athletes in Competition in Term of sudden trauma.
Some doubts remain about the possible danger of the complementary movements carried out to perfect the result of the technique in competition, as previously indicated.

## 11. Discussion

The experimental part of this research applied to the safety of Suwari Seoi Family against the immediate trauma, was performed at Italian Olympic Center "Matteo Pellicone" in Rome with 8 children ranging from Es. Category A till to junior: 5 Male and 3 Female, and 18 Athletes ranging from Junior to Senior: 12 Male and 6 Female.

There are evaluated two form of Suwari Throws, one with feet pointed in which Tori is in deep kneeling position, and another one in which Tori is in "seiza" position.

In the experimental protocol was described the experiment and the lag of time used for safety ( 3 days for Children, 2 days for Athletes)

The first interesting results was that for children, the speed of knees dropping down was decreasing from alone till to throwing Uke.

This means in term of safety that the presence of a body to throws slows down the speed of fall, making the knees' impact on the Tatami less risky for the knee joint.

For the Adult both male and female show one interesting same behavior, 60% of subjects slowed down, 30% accelerated and 10% showed no increase or decrease in speed.

This almost standard behavior could be connected to the personal skill of athletes in perform this technique.

The Specific calculation of the compression of the tatami during the trials both for children and Athletes, on IJF licensed Tatami in PU range between 0,88 mm till to 2,28mm

Depending obviously from the mass and velocity involved.

It is well known in biomechanics and orthopedic world that the knee structure can absorb part of stress moving forward-backward of about 5-10 mm [64].

Absurdly even in the case in which we no consider the decreasing stress received by knees (more or less15- 20% less of the stress given to the Tatami on the basis of the collision laws) but the total stress received by tatami, taking into account the elastic properties of the knee, anyway the final results, assure that if the technique is performed with no mistakes, on a IJF licensed Tatami, the impact received will be of negligible effect with respect to the target organ the Knee : the Posterior Cruciate Ligament ( PCL).

## 12. Conclusions

Even we consider the worst stress on the knees or knees and tibia ranging: from 0,11 MPa till to 0,22 MPa. For a 12 cm thick knee in the sagittal plane, the relative strain, considering the young modulus of tibia bone 18,1 GPa [67], gives a very small compression which provides as result a negligible stress on the cruciate posterior ligament, who is able to resist at a force of 1500 N [68]. The final result, despite this small number of samples, assures that *if the technique is performed with no mistakes, on a IJF licensed Tatami, the impact received will be of negligible effect with respect, for sudden trauma, to the target ligament of the Knee : the Posterior Cruciate Ligament (PCL).*

A German paper on 260 athletes at a national level that deals with traumas related to techniques including those of the knee [15] shows that for Seoi, our target structure, PCL is connected to 0.05%, whereas ACL and Patella to 0.15% , MCL and LCL to 0.10% , then because the others knee's target structures are much more connected with Seoi than PCL , it is possible to state that : *the deduction derived from the theoretical analysis carried out, indicating that, probably, most of the damage, that occur to the different knee ligaments, are almost certainly produced, during the complementary movements, made to refine the result in competition, it's right.*

Although surely related both: to low number of samples and to safety assumption made on the impact forces, the experimental results on standard children show that a slightly larger stress is associated with the style of feet stretched towards pointed feet, males (0.17 vs 0.11) MPa; females (0.22 vs 0.13) MPa.

In male Standard Adult Athletes this tendency seems to be slightly reversed (0.12 vs 0.14) MPa, while in female Standard Adult Athlete it appears to be conserved (0.13 vs 0.12) MPa.

However, it is more correct to say that for both standard adult athletes it seems that there are no substantial differences regarding their safety between the two performance styles analyzed.

As already underlined in this paper we start with this new methodology, applied for the first time on Tori safety. Indeed, the result of the proven methodology in use, is limited by the small number of samples analyzed. Of course, more data will be collected, best estimation will be made to evaluate the standard athletes**,** and consequently, to obtain more precise evaluations about the safety of the technique.

In light of the above, this work must be considered substantially, as the right master working track to conduct more and even more precise researches, on the Tori's safety in Suwari Seoi execution

*13. References*
1. **Hidetoshi Nakanishi** *"Seoi nage". Pag. 9-10 Publisher Ippon Books 1992 ISBN 0-9518455-4-3*
2. **Toshiro Daigo** *"Kodokan Judo Throwing techniques" pag 24 Kodansha International 2005 ISBN 4 -7700-2330-8*
3. **Kazuzo Kudo** *"Dynamic Judo Throwing techniques" pag. 220 Japan Publication Trading Co, Ltd 1966 LCC card number 66-21211*
4. **Gunji Koizumi** *" My study of judo" pag 51 Sterling publishing Co. Inc. 1960 LCC card number 61-15858*
5. **Thierry Paillard**; *"Optimization de la Performance Sportive en Judo". De boeck , 2010, ISBN-13: 978-1-4612-8733-9*
6. **Siliski** *"Traumatic knee disorders" Springer 1994 ISBN-978-2-8041-0783-3*
7. **Fowler PJ, Messieh SS.** *"Isolated posterior cruciate ligament injuries in athletes". Am J Sports Med. 1987; 15:553-557.*
8. **Rich and Coworkers** *"Forensic Medicine Of The Lower Extremity Human Identification And Trauma Analysis Of The Thigh, Leg, And Foot " Humana Press 2005 ISBN 1-58829-269-X*
9. **Koshida and Coworkers** *"The common mechanisms of anterior cruciate ligament injuries in judo: a retrospective analysis" Br J Sports Med 2010 44: 856-861 originally published online November 28, 2008.*
10. **Han and Coworkers** *"Gender Differences in Lower Extremity Kinematics During High Range of Motion Activities Journal of Medical Imaging and Health Informatics Vol. 4, 272–276, 2014*
11. **Murray and Coworkers** *"The ACL Handbook" Springer 2013 ISBN 978-1-4614-0759-1*
12. **Carretero & Lopez Elvira** *:" Impacto producido por la tecnica Seoi Otoshi . Relacion con anos de practica y grado de judo" RAMA vol.9 , 2014 .*
13. **C. Torres** Thesis Doctoral *" El Seoi-Nage De Rodillas Incorrecta Para La Enseñanza En Niños Y Adolescenetcecsió. Sne Toéic-Nnaicgae De Pie Técnica De Elección" Universidad Las Palmas de gran Canaria 2015*
14. **Barsottini & coworkers** *"Relationship between techniques and injuries among judo practitioners" Rev Bras Med Esporte _ Vol. 12, Nº 1 – Jan/Fev, 2006*
15. **Prill & Alfuth** *"the influence of the special throwing techniques on the prevalence of knee joint injuries " Archives of Budo September 2014*
16. **Handler & coworkers** *"Technical-tactical preparation of Austrian judoka at the Austrian national championships and the number of associated injuries" preprint DOI* [10.17605/OSF.IO/Q23TV](10.17605/OSF.IO/Q23TV)
17. **Du Bois D, Du Bois EF** *"A formula to estimate the approximate surface area if height and weight be known". Archives of Internal Medicine.* **17** *(6): 863–71, Jun 1916.*
18. **Sacripanti** *"The road to Ippon" Zagreb 2015* [https://arxiv.org/abs/1506.01812](https://arxiv.org/abs/1506.01812);
19. **Calderon, Gil;** *"Experimentos con objectos que caen con aeleration major que g" Lat. Am. J. Phys. Educ. Vol. 5, No. 2, June 2011*
20. **Erman ;***"Faster than Gravity" Seminar Ljiublijana 2007* [http://mafija.fmf.uni-lj.si/seminar/files/2007_2008/Faster_than_gravity.pdf](http://mafija.fmf.uni-lj.si/seminar/files/2007_2008/Faster_than_gravity.pdf);
21. **Vareschi, Kamiya** *"Toy models for the falling chimney"* [https://arxiv.org/pdf/physics/0210033.pdf](https://arxiv.org/pdf/physics/0210033.pdf);

22. ***Laurent Blais,*** Thesis *"Analyse objective de deux techniques de projection en judo : seoï nage et uchi mata de la réalité mécanique aux applications pédagogiques". Poitiers University 2004*
23. ***L. Blais , Trilles , Lacoture*** *Three-dimensional joint dynamics and energy expenditure during the execution of a judo throwing technique (Morote Seoï Nage) Journal of Sport science 2007*
24. **Ishii and Ae** : *"Biomechanical factor of effective Seoi Nage in Judo" Doctoral program in Physical Education Fitness and Sport Science Tsukuba Japan 2014*
25. **Ishii and Coworkers:** *"Front turn movement in Seoi Nage of elite Judo Athletes" 30 annual conference of Biomechanics in sport Melbourne 2012*
26. **Ishii and Coworkers:** *The Centre Of Mass Kinematics For Elite Women Judo Athletes In Seot-Nage " 34 annual conference of Biomechanics in sport Tsukuba Japan 2016*
27. **Ishii and Ae*:*** *Comparison Of Kinetics Of The Leg Joints In Seoi-Nage Between Elite And College Judo Athletes " 33 annual conference of Biomechanics in sport Poitiers France 2015*
28. **Ishii and Coworkers:** *Comparison Of Angular Factors To Determine Quickness In Seoi-Nage Between Elite And College Judo Athletes 2011*
29. **M. Ikai & Y. Matsumoto***: "The Kinetic of judo" Bulletin of the Association for the scientific studies on judo, kodokan report I 1958 Tokyo, Japan.*
30. **Hassmann and coworkers** : *"Judo performance test using a pulling force device simulating a Seoi Nage throws" journal of martial art anthropology N° 3, 2011*
31. **Lopes Melo and Coworkers** : *"Cinematica da variacao angular de tronco, quadril e johelo do atacante na tecnica Seoi Nage no Judo" Universidad. de Santa Catarina Florianopolis Brazil 2004*
32. **Franchini and coworkers** : *"Energy expenditure in different judo throwing techniques" Proceeding of first joint international pre-olympic conference of sports science and sports engineering Nanjing China 2008*
33. **Ibrahim Fawzi Mustafa**, *"Force impulse of body parts as function for prediction of total impulse and performance point of Ippon Seoi Nage skill in judo" World Journal of Sport science N° 3 , 2010*
34. **Blais, Trilles** : *"Analyse méchanique comparative d'une meme projection de Judo: Seoi Nage, realisée par cinq experts de la Fédération Française de Judo". Science et Motricité N°51, 2004*
35. **Blais, Trilles, Lacoture** : *"Détermination des forces de traction lors de l'exécution de Morote Seoï Nage réalisée par 2 experts avec l'ergomètre de Mayeur et un partenaire " Journal of Sport Science 2007*
36. **Thiers and Coworkers** : *"Sport analysis using Shimmer ™ sensors" German Journal for young researchers 2012*
37. **Sacripanti and Coworkers***: "Infrared Thermography-Calorimetric Quantitation of Energy Expenditure in Biomechanically Different Types of Jūdō Throwing Techniques. A Pilot Study". Annals of Sport medicine and researches 2015*
38. **Aoki and Coworkers:** *"Biomechanical analysis of Seoi Nage in judo throwing techniques" In Japanese 1986*
39. **Ji Tae & Seong- Gyu** *:" A kinematic analysis of Morote Seoi Nage according to the Kumi Kata types in Judo" Korean Journal of Sport Biomechanics Vol 16 N°2 2006*

40. **Chwala, Ambrozy, Sterkowicz** : *"Tridimensional analysis of the jujitsu competitors motion during the performance of the Ippon Seoi Nage throw" Archive of Budo science of martial art and extreme sports N°9  2013*
41. **Peixoto, Monteiro** : *"Structural Analysis And Energetic Comparison Between Two Throwing Tecniques  Uchi-Mata Vs Ippon-Seoi-Nage". 3 rd  European Science of Judo Symposium - 26 April 2012 Chelyabinsk*
42. **Gutierrez- Santiago & Coworkers**: *" Sequence of error in the Judo Throw Morote Seoi Nage and their relationship to the learning process" Journal of Sport Engineering and technologies  2013*
43. **Imamura and Coworkers:** *"A three-dimensional analysis of the center of mass for three different judo throwing techniques" Journal of Sport Science and Medicine 2006*
44. **Gaskell and Laughlin** *"Introduction to the Thermodynamics of Materials" CRC Press 2018  ISBN 9781498757003*
45. **Ionescu M**. *Chemistry and Technology of Polyols for Polyurethanes, 2nd Edition Volume I&II Smithers Rapra Publisher 2016 ISBN: 978-1-91024-213-1*
46. **Mane & coworkers** *Mechanical Property Evaluation of Polyurethane Foam under quasi-static and Dynamic Strain Rates- An Experimental Study Procedia Engineering 173 ( 2017) 726 – 73*
47. **Bradley & Sullivan** *Thermal Expansion of Polyurethane Foam 43rd Annual Technical Meeting of the Society of Engineering Science The Pennsylvania State University August 2006*
48. **Sychev** *Complex Thermodynamic Systems (Studies in Soviet Science) Consultant bureau New York 1973.*
49. **Jou and Coworker** *"Extended Irreversible Thermodynamics" Springer 2010 ISBN 978-90-481-3073-3*
50. **Sacripanti** *" A Seoi survey for coaches and teachers"* [https://arxiv.org/ftp/arxiv/papers/1506/1506.01372.pdf](https://arxiv.org/ftp/arxiv/papers/1506/1506.01372.pdf)
51. **Sacripanti** *"Judo Biomechanical Science for IJF Academy"  In Print  2018*
52. **Hirokawa & Fukunaga** *" Knee Joint Forces when rising from kneeling positions" Journal of Biomechanical science and engineering  Vol 8  N 1 -2013*
53. **Fukunaga & Morimoto** *"Calculation of the knee joint force at deep squatting and kneeling" Journal of Biomechanical science and engineering  Vol 10  N 4 -2015*
54. **Nagura & Coworkers** *"Mechanical loads at the knee joint during deep flexion" Journal of Orthopaedic Research 20; 881- 886; - 2002*
55. **Bisciotti and Coworkers** *"Analisi delle caratteristiche elastiche dell'unità muscolo tendinea e delle capacità di equilibrio di due diverse tipologie atletiche" Medicina dello Sport , 53-2-200*
56. **Lévesque** *"Law of cooling, heat conduction and Stefan-Boltzmann radiation laws fitted to experimental data for bones irradiated by $CO_2$ laser"* [Biomed Opt Express](). *2014 Mar 1; 5(3): 701–712.*
57. **Latif**  *"Heat Conduction"  Springer 2009  ISBN  978-3-642-01266-2*
58. **Quesada** *"Application of infrared thermography in Sport Science" Springer 2017 ISBN 978-3-319-47409-0*
59. **Sacripanti & coworkers** *"Infrared Thermography- Calorimetric Quantitation of Energy Expenditure in Biomechanically Different Types of Jūdō Throwing Techniques. A Pilot Study" Annals of Sport Medicine and Research 2015*
60. **Springs** [http://www.edutecnica.it/meccanica/molla/molla.htm](http://www.edutecnica.it/meccanica/molla/molla.htm),

61. **Thierry Loison** *"Personal Communication  03-28-2018"*
62. **Costeff H**, *"A simple empirical formula for calculating approximate surface area in children.," Arch Dis Child, vol. 41, no. 220, pp. 681–683, Dec. 1966.*
63. **Sacripanti, De Blasis** *"Safety for Judo children : Methodology and Results", June 2017* [https://arxiv.org/abs/1706.05627](https://arxiv.org/abs/1706.05627).
64. **Purvi SD Patel and Coworkers** *"Compressive properties of commercially available polyurethane foams as mechanical models for osteoporotic human cancellous bone"* 2012 [https://www.ncbi.nlm.nih.gov/pmc/articles/PMC2575212/](https://www.ncbi.nlm.nih.gov/pmc/articles/PMC2575212/)
65. **Lee** *"Biomechanics of Hyperflexion and Kneeling before and after Total Knee Arthroplasty"* 2014 [https://www.ncbi.nlm.nih.gov/pmc/articles/PMC4040370/](https://www.ncbi.nlm.nih.gov/pmc/articles/PMC4040370/)
66. **Yoshitaka, Y.,** *JUDO waza no daihyakka ( Encyclopaedia of JUDO techniques) (in Japanese). Vol. 1 Baseball Magazine sha,. (2015) ISBN978-4-583-10794-3.*
67. **AAVV** *" Materiali Biologici" La Sapienza University  2010* [http://dma.ing.uniroma1.it/users/scicostr_c1/RdBM_Cap_4.pdf](http://dma.ing.uniroma1.it/users/scicostr_c1/RdBM_Cap_4.pdf)
68. **A. Race & A. Amis** *"The mechanical properties of the two bundles of the human posterior cruciate ligaments" Journal of Biomechanics Vol. 27. No. I. pp 13 24, 1994*